\documentclass[12pt,draftcls,onecolumn]{IEEEtran}% Aptara syntax

\IEEEoverridecommandlockouts

\usepackage[ruled]{algorithm2e}

\SetAlFnt{\small}
\SetAlCapFnt{\small}
\SetAlCapNameFnt{\small}
\SetAlCapHSkip{0pt}
\IncMargin{-\parindent}

\usepackage{amsfonts, amsmath, amssymb}
\usepackage{epsfig,graphicx,subfigure}

\newtheorem{assumption}{Assumption}
\newtheorem{definition}{Definition}

\newtheorem{proposition}{Proposition}
\newtheorem{observation}{Observation}
\newtheorem{corollary}{Corollary}
\newtheorem{theorem}{Theorem}
\newtheorem{lemma}{Lemma}

\newcommand{\nullvendor}{s_{\bot}}
\newcommand{\sellerset}{\mathcal{S}}
\newcommand{\sellersetc}{\sellerset^c}
\newcommand{\buyerset}{\mathcal{B}}
\newcommand{\market}{\mathcal{M}}
\newcommand{\network}{\mathcal{G}}
\newcommand{\alg}{\mathcal{A}}
\newcommand{\fflows}{\mathcal{F(\mathcal{N})}}
\newcommand{\rtrans}{\mathcal{T(\mathcal{M})}}
\newcommand{\feasible}{\Pi}
\newcommand{\C}{\mathcal{C}}
\newcommand{\X}{\mathcal{X}}
\newcommand{\K}{\mathcal{K}}
\newcommand{\N}{\mathbb{N}}
\newcommand{\RR}{\mathbb{R}^+}
\newcommand{\Tau}{\mathcal{T}}
\newcommand{\bigO}{\mathcal{O}}
\newcommand{\bigTheta}{\Theta}
\newcommand{\bigOmega}{\Omega}

\newcommand{\vAB}[3]{v_{#1}(#2,#3)}
\newcommand{\tra}[2]{t_{#1\rightarrow #2}}
\newcommand{\gtra}[3]{\bar t_{#1\rightarrow #2#3}}
\newcommand{\gtraset}[2]{\bar t_{#1\rightarrow #2}}
\newcommand{\posi}[1]{\mathcal{P}(#1)}
\newcommand{\nega}[2]{\mathcal{N}(#1,#2)}
\newcommand{\negaset}[1]{\mathcal{N}(#1)}

\newcommand{\tposi}[1]{P(#1)}

\newcommand{\tnegaset}[1]{N(#1)}

\newcommand{\mktpricex}[3]{p_{#1}^{#2}(#3)}
\newcommand{\mktpriceb}[2]{p_{#1}^{#2}}

\usepackage{color}

\begin{document}

% Title portion
\title{Group buying with bundle discounts:\\computing efficient, stable and fair solutions}
\author{Lorenzo Coviello, Yiling Chen, Massimo Franceschetti\thanks{LC: Media Lab, Massachusetts Institute of Technology. YC: School of Engineering and Applied Science, Harvard. MF: Department of Electrical and Computer Engineering, University of California San Diego. Email:
        {\tt\small lorenzoc@mit.edu}}}
\date{}

\maketitle

\begin{abstract}
We model a market in which nonstrategic vendors sell items of different types and offer bundles at discounted prices triggered by demand volumes. Each buyer acts strategically in order to maximize her utility, given by the difference between product valuation and price paid.
Buyers report their valuations in terms of prices on sets of items, and might be willing to pay higher prices in order to subsidize other buyers and to trigger discounts.
The resulting price discrimination can be interpreted as a redistribution of the total discount.
We consider a notion of stability that looks at unilateral deviations, and
show that efficient allocations -- the ones maximizing the social welfare -- can be stabilized by prices that enjoy desirable properties of \emph{rationality} and \emph{fairness}.
These dictate that buyers pay higher prices only to subsidize others who contribute to the activation of the desired discounts, and that they pay premiums over the discounted price proportionally to their surplus -- the difference between their current utility and the utility of their best alternative.
Therefore, the resulting price discrimination appears to be desirable to buyers.
Building on this existence result, and letting $N$, $M$ and $c$ be the numbers of buyers, vendors and product types, we propose a $\bigO(N^2+NM^c)$ algorithm that, given an efficient allocation, computes prices that are rational and fair and that stabilize the market.
The algorithm first determines the redistribution of the discount between \emph{groups} of buyers with an equal product choice, and then computes single buyers' prices.
Our results show that if a desirable form of price discrimination is implemented then social efficiency and stability can coexists in a market presenting subtle externalities, and computing individual prices from vendor prices is tractable.
\end{abstract}

\section{Introduction}
\label{market:sec:intro}
We model a market in which vendors offer items of different types, and each buyer is interested in purchasing a unit of each type, possibly from different vendors.
Vendors are nonstrategic, supplies are unlimited and each vendor has a price for each item.
Moreover, each vendor has a discount schedule according to which the bundle of all items is offered at discounted price if her demands exceed given thresholds.
In this work, we assume that only buyers who buy all product types from a single vendor (i.e., a bundle) can enjoy a discounted price.
This can be seen as an incentive to loyal customers who buy from a single vendor who can sustain lower sale prices only in an economy of scale.
Buyers play strategically and each selfishly tries to maximize her utility, given by the difference between the perceived value of the products and the price paid.
A buyer might have different valuations of a given product type offered by two different vendors (as in the case of TVs of different brands).
Buyers report their valuations in terms of prices on sets of items, and might be willing to pay higher prices in order to subsidize other buyers and to trigger discounts.
Buyers who do not purchase any bundle (i.e., who buy from several vendors) also contribute to the activation of discounts by increasing demand.
The resulting price discrimination can be interpreted as a redistribution of the total discount, a form of cooperation between buyers.
In general, price discrimination is known to increase market efficiency and to allow optimal resource allocation~\cite{phlips1983economics, shapiro2013information, varian1989price, varian1996differential}, and we investigate it in a context presenting subtle externalities between buyers' choices.
To illustrate the potential benefit of the proposed pricing scheme, consider a buyer who desires the bundle from a certain vendor at a discounted price.
In order to trigger the discount, she might be willing to pay a higher price in order to subsidize other buyers and induce them to purchase from the same vendor (as they would otherwise prefer other vendors).

%The externalities present in this scenario (for which a buyer's utility depends on the choices of others) make the core of the game empty in general~\cite{osborne}.
%That is, given a market, there might be no configuration such that no coalition of buyers can increase their total utility by deviation.
We focus our attention on market configurations, or \emph{allocations}, that maximize the social welfare.
These cannot be sustained in general, as buyers might prefer to deviate, and therefore we assume that buyers can pay prices that are different between each other.
Given an allocation and prices, we introduce a notion of stability that looks at unilateral deviations by buyers.
%The proposed notion of stability is suitable for a setting in which prices are determined given buyers' choices after a deal is over, and therefore buyers cannot enjoy discounts by signing up to deals after deviation.
The proposed notion of stability is suitable for a setting in which prices are determined by buyers' choices and buyers cannot enjoy discounts after deviation.
This assumption is in line with common group buying platforms, where buyers are given limited time to sign up to a discount offer.
Moreover, we assume a scenario where communication and coordination between buyers is mediated by a central entity (e.g., an online setting in which buyers are allowed to set  the maximum prices they are willing to pay for given sets of items).
Given a market configuration, we ask whether there exist prices such that no buyer wants to deviate.

The proposed pricing scheme models an on-line market in which buyers indicate their willingness to pay for products from different vendors, which in turn offer discounts if enough people sign up.
As such, it is related to ``deal-of-the-day'' on-line purchasing, made popular by Groupon and Living Social, in which sellers offer discounted gift certificates that become valid if enough people sign up to the deal.
In addition, in on-line shopping platforms such as Ebay, buyers specify the maximum amount they are willing to pay for a product.
According to our model, each buyer prefers the set of products with the higher difference between her valuation and the price, and buyers might be willing to pay prices that are different between each other (i.e., redistribute the total discount) in order to trigger deals.
Even if the selling price of a product is higher than her valuation, a buyer might be willing to purchase it if offered a lower personalized price (i.e., somebody else bears part of the cost).
Similarly, if a buyer's valuation of a product choice is high enough with respect to a discounted price, then she might be willing to pay a higher price to decrease the effective price of other buyers and induce them to buy  -- contributing to the activation of the discount.
In such a scenario, a stable assignment of buyers to vendors and the final buyers' prices must be computed in a centralized fashion.

Given an allocation that maximizes social welfare, it is trivial to prove the existence of prices such that no buyer is unilaterally willing to deviate from her current product choice.
However, arbitrary prices might be undesirable for buyers,
and we look at prices that enjoy additional properties of \emph{rationality} and \emph{fairness}.
Rationality dictates that only buyers who benefit from discounts might pay higher prices to subsidize other buyers, and only in order to subsidizes buyers who purchase (at least one item) from the same vendor (as they might be necessary to trigger the discount).
This is motivated by the willingness of each buyer to subsidize only buyers she benefits from.
Fairness dictates that buyers pay premiums over the discounted price proportionally to their \emph{surplus}, that is, the difference between their current utility and the utility of their best alternative.
In order to motivate this notion of fairness, observe that it might be undesirable for a buyer to pay a disproportionately large amount of the subsidy needed by the buyers she benefits from if there are other buyers willing to contribute
(although, from the strict point of view of stability, a buyer might be willing to give up her entire surplus regardless of the behavior of others).

\paragraph*{Summary of results}

In Section~\ref{market:sec:model}, after introducing the model,
we establish a correspondence between buyers' prices and transfers of utility between buyers.
In Section~\ref{market:sec:existence}, we show that, given any allocation that maximizes the social welfare (or SWM allocation), there exist rational prices that stabilize it (Theorem~\ref{thm:existence} and Corollary~\ref{cor:existence} in Section~\ref{market:sec:existence}).
This means that \emph{efficient} allocations are also \emph{stable} up to suitable prices (the \emph{price of stability} is one, a property that is not always observed in games~\cite{jackson,roughgarden}).
To prove this, we partition buyers according to their choices and surplus: on one side, groups of ``rich'' buyers getting the same discounted bundle and with a positive surplus (i.e., willing to pay higher prices in order to subsidize other buyers); on the other side, groups of ``poor'' buyers with the same product choice and negative surplus (i.e., in need of additional discount).
Then, we show that there are ``rational'' transfers \emph{between groups of buyers} such that:
each rich group subsidize poor groups with at least one vendor in common;
each rich group transfer at most their available surplus;
and each poor group receive the necessary subsidy.
Group transfers can be interpreted as a redistribution of discounts between groups, and can be translated into rational and stabilizing buyers' prices.
This existence result constitutes the main contribution of this work, and its proof is based on the construction of a graph which encodes the transfers between groups of buyers and has no edges if and only if the property of rationality is satisfied.

In Section~\ref{sec:computation_transf}, we show how prices that are rational and fair and stabilize the market can be efficiently computed given a SWM allocation.
First, transfers between group of buyers are computed via the Ford-Fulkerson algorithm for the maximum flow on a network such that rational group transfers and feasible flows are in one-to-one correspondence (Section~\ref{sec:computation_transf_group}).
Then, rational and fair prices who stabilize the SWM allocation are computed (Section~\ref{sec:computation_transf_buyers}).
For a market with $N$ buyers, $M$ vendors and $c$ product types, group transfers are computed in time $\bigO(M^cT)$,\footnote{Consider two functions $f(x)$ and $g(x)$ of a vector $x=(x_1,\ldots,x_n)$. We say that $f(x)=\bigO(g(x))$ if there exist constants $C>0,m$ such that $f(x)\le C g(x)$ for all $x$ such that $\min_{1\le i\le n}x_i\ge m$.
We say that $f(x)=\bigOmega(g(x))$ if there exist constants $C>0,m$ such that $f(x)\ge C g(x)$ for all $x$ such that $\min_{1\le i\le n}x_i\ge m$.
We say that $f(x)=\bigTheta(g(x))$ if both $f(x)=\bigO(g(x))$ and $f(x)=\bigOmega(g(x))$.}
where $T$ is the total subsidy needed.
If prices and buyers' valuations do not depend on $N$ and $M$, this is $\bigO(NM^c)$.
Buyers' prices are computed from group transfers in additional time $\bigO(N^2+NM^{c-1})$, for a cumulative time of $\bigO(N^2+NM^c)$.
If the number of vendors is constant (or grows as $M^c=\bigO(N)$) then the overall complexity is dominated by the $N^2$ term.

Section~\ref{sec:swm} deals with the computation of SWM allocations.
A natural approach consists in a mixed integer program, see~\cite{rothblum}, requiring time exponential in $N$ and $M$, and whose relaxation is not guaranteed to have integral solutions (i.e., corresponding to valid allocations).
Conditional on the number of buyers assigned to each pair of vendors, we compute a SWM allocation in time $\bigTheta(N^2M^c)$ via the Ford-Fulkerson algorithm for the maximum flow with minimum cost on a network such that maximum flows and feasible allocations are in one-to-one correspondence.
Computing a SWM allocation requires to consider a number of cases of the order of $N^{M^c}$, and this term dominates the computational complexity.
Getting rid of the exponential dependency on $M$ does not seem possible, due to the theoretical hardness of the problem.
However, the overall time complexity is polynomial in $N$, and usually $M$ can be assumed much smaller than $N$ or even constant.

\paragraph*{Related work}

Online retailing has seen a continuous growth during the last two decades~\cite{laudon}.
``Deal of the day'' websites such as Groupon and Living Social have introduced a new form of buying, in which enough buyers must sign up for a deal to be valid.
An overview of the literature on group buying in the web is given by Anand and Aron~\cite{anand} and by Kauffman and Walden~\cite{kauffman}.
Cai et al.~\cite{cai2013designing} frame the market design problem associated to daily deals.
Chen et al.~\cite{chen2010segmenting} show that group buying is more effective when low-valuation demand is larger than high-valuation demand.
Matsuo et al.~\cite{matsuo2005volume} and Prashanth and Narahari~\cite{prashanth2008efficient} proposed efficient allocation mechanisms for group buying in the presence of volume discounts.
Yamamoto and Sycara~\cite{yamamoto2001stable} and Li et al.~\cite{li2005mechanism} proposed mechanisms for coalition formation in the context of group buying, based on transferable utility.

The present paper is related to an established body of literature on price discrimination, a practice that has become increasingly popular after the advent of online markets~\cite{laudon}.
The pricing mechanism we propose consists in two levels of price discrimination.
On the one hand, sellers offer discounted prices only on bundles of products.
Bundling has long been used as a technique to implement and conceal price discrimination, determined by buyers' choice and therefore not resulting in privacy loss~\cite{odlyzko2003privacy}.
On the other hand, final buyers' prices result from a possibly unequal redistribution of the total discount, according to which buyers might pay different prices even if they buy the same products.
This redistribution of the discount is what allows social efficient allocations to be maintained.
In general, price discrimination is considered a desirable practice that increases market efficiency and allows optimal resource allocation~\cite{phlips1983economics, varian1989price, varian1996differential, shapiro2013information}.
The simple idea at its basis is to charge higher prices to buyers who are willing to pay more (as they have a higher valuation of the products).
Even if buyers might be in general reluctant to disclose their valuations, such information can be inferred via technology that is standard in online retailing, often at the expense of individual privacy~\cite{odlyzko2003privacy}.
For example, an individual's willingness to pay can be inferred from her purchase history or from the attributes she might choose to disclose~\cite{acquisti2005conditioning}.
Price discrimination online can be detected~\cite{mikians2012detecting}, and raised concern in some case~\cite{heffernan2010amazon}.
A review of the literature on behavior-based price discrimination is provided by Fudenberg and Villas-Boas~\cite{fudenberg2006behavior}.
Buyers' reluctancy to disclose personal information and other concerns with price discrimination
might be overcome by the guarantee of reaching more efficient resource allocations.
In particular, we focus on a form of price discrimination (through a redistribution of the available discount) that satisfies a desirable property of rationality, according to which a buyer is willing to give up part of the discount she deserves only in order to support buyers whose choice she benefits from.

When externalities are present in the market, stability often becomes problematic~\cite{hatflieldmilgrom,sasaki}.
In the context of group buying, the externalities are the numbers of buyers purchasing each product from each vendor, as they determine who benefits from discounts.
%That is, given a market, there might be no configuration such that no coalition of buyers can increase their total utility by deviation
Due to the interdependencies between buyers' choices and utility, only specific assumptions can guarantee the existence of configurations such that no coalition of buyers can increase their total utility by deviation~\cite{osborne}.
This is the case when buyers have no preferences over vendors and there are no discounts (considered by Huevel et al.~\cite{huevel}), or when preferences derive from transportation costs and either there are no discounts or there are discounts but all buyers buy from the same vendor (considered by Chen~\cite{chen}).
In general situations, when discounts and preferences are both present, notions of stability that consider unilateral deviation are considered.

Our model is closely related to Lu and Boutilier~\cite{luboutlier}, in which multiple vendors sell a single type of product and post discounts at increasing demand volumes.
They consider markets with transferable and non-transferable utility, showing that several notions of stability (looking at unilateral deviation) can be guaranteed on both. In particular, in the case of transferable utility, stability and efficiency coexist.
Our work builds on the model by Lu and Boutilier~\cite{luboutlier}, and the novelty of our contribution is twofold.
First, we extend the model to a more general case of multiple products on the market and to the possibility for vendors to activate discount on bundles of items rather then single items.
As discounts can be triggered by buyers who do not necessarily benefit from them, proving the existence of (rational) prices that stabilize the market is nontrivial, and it is accomplished in the present work through a graph theoretical argument (see Section~\ref{market:sec:existence}).
Second, we also consider the computational side of stability, by proposing a simple and efficient algorithm to compute prices that stabilize the market and enjoy desirable properties of \emph{rationality} and \emph{fairness}.
In particular, fairness of buyers' prices dictates that no buyer pays disproportionally large subsidies.
Our model, results and algorithms include the single-item market by Lu and Boutilier~\cite{luboutlier} as a special case.

In the present work, we only consider the point of view of buyers and assume that pricing schedules of vendors are fixed.
Research on supply chain management suggests that posting volume discounts increases efficiency from the vendors' point of view~\cite{lal1984approach,monahan1984quantity}.
In the context of group buying, Meir et al.~\cite{meir2014value} considered the point of view of vendors and showed that in general it might be preferable for a vendor to post discounts if other vendors post discounts.
Edelman et al.~\cite{edelman2011groupon} characterize the effect of discounts on customer acquisition.

Our work is also related to an established line of research on matching models and allocation mechanisms.
Such models have received considerable attention by computer scientists~\cite{israeli2010,kanoria2011fast,pettie2004simpler,rothblum} and economists~\cite{ashlagi2011ec,galeshapley,hatflieldkominers,hatflieldmilgrom,rothsotomayor,roth}, as they constitute the abstraction of real world strategic scenarios such as retail markets, the labor market, college admissions, and the assignment of residents to hospitals.

\section{The model}
\label{market:sec:model}
We consider a market $\market$ consisting of a set of $N$ buyers $\buyerset$ and a set of $M$ vendors $\sellerset$. Each vendor sells items (or products) of $c$ types denoted by $1,\ldots,c$, and we assume supplies are unlimited. Let $C=\{1,\ldots,c\}$.
Each buyer is willing to purchase a single unit of each item type, possibly from two or more different vendors\footnote{We do not make any assumption about the nature of the products, which are not assumed to be complements or substitutes. We only assume that buyers assign zero or negative valuation to sets of products involving multiple units of any single item.}.
As a remark, a vendor $\nullvendor\in\sellerset$, called the \emph{null vendor}, might represent the choice not to buy (i.e., buyer $b$ choosing item $k$ from $\nullvendor$ means that $b$ does not buy item $k$). In what follows, prices, discounts and valuations corresponding to such vendor $\nullvendor$ will be pointed out.
Let $\sellersetc$ denote the cartesian product of $c$ copies of $\sellerset$.

An allocation is a set of tuples $\mu\subset\buyerset \times \sellersetc$ such that each $b\in\buyerset$ appears in \emph{exactly} a single tuple.
An allocation represents buyers' choices and, for $\bar s =(s_1,s_2,\ldots,s_c)\in\sellersetc $, $(b,\bar s)\in\mu$ denotes that $b\in\buyerset$ purchases item $k$ from vendor $s_k$ for $k=1,\ldots,c$.
We write $\mu(b)=\bar s$, and $\mu^k(b)=s_k$ for $k\in C$.

Given an allocation $\mu$, for each $\bar s\in \sellersetc$, let $\hat\mu(\bar s)=\{b\in\buyerset:\mu(b)=\bar s\}$ be the set of buyers who purchase item $k$ from vendor $s_k$ for all $k\in C$, and let $n(\bar s)=\vert\hat\mu(\bar s)\vert$ be its cardinality.
Given a allocation $\mu$, for each $s\in \sellerset$ and $k\in C$, let $\hat\mu^k(s)=\{b\in\buyerset:\mu^k(b)=s\}$ be the set of buyers who purchase item $k$ from vendor $s$ and $n^k(s)=\vert\hat\mu^A(s)\vert$.
We refer to $n(s)=(n^1(s),\ldots,n^c(s))\in\N^c$ as the \emph{demand vector} of vendor $s$ (where $\N$ is the set of nonnegative integers).

\paragraph*{Price schedules}
Vendors are nonstrategic. The prices offered by a vendor are determined by her demand vector, according to a \emph{price schedule} defined as follows.
%Let $\C=\{x\subseteq C\}$ be the partition of $C$ (i.e., the set of all $2^c$ subsets of $C$).
%The price schedule $p_s$ of vendor $s\in\sellerset$ is a mapping from $\N^c\times\C$ to $\R$ (the set of nonnegative real numbers), such that, for $n\in\N^c$ and $x\in\C$, $p_s(n,x)$ is the price for the bundle of products $x$ offered by $s$ under demand $n$. Let $p_s(n,\emptyset)=0$ for each $s$ and $n$, and $p_{\nullvendor}(n,x)=0$ for all $n$ and $x$.
%We require that $p_s(m,x)\le p_s(n,x)$ for all $x\in\C$ if $m\ge n$ component-wise.
%Letting $e_k$ be the unit vector with the $k$-th component equal to one and all other components equal to zero, we refer to $p_s^k=p_s(e_k,\{k\})$ as the \emph{base price} of item $k$ offered by $s$.
Each vendor $s\in\sellerset$ has a base price $p_s^k$ for each item $k\in C$ (we let $p_{\nullvendor}^k=0$ for the null vendor $\nullvendor$).
Moreover,  $s$ activates discounted prices on the bundle of all items $C$ when certain thresholds are met, as we explain next.
Let $p_s^{(0)}=\sum_{k\in C}p_s^k$ the \emph{base price} of all items offered by $s$.
We assume that $s$ has $h$ vectors $\tau_i(s) = (\tau^1_i(s),\ldots,\tau^c_i(s))$ for $i=1\ldots h$, called the demand thresholds vectors of $s$, such that $\tau^k_{i+1}(s)\ge \tau^k_i(s)$ and $\sum_{k\in C}\tau^k_{i+1}(s)> \sum_{k\in C} \tau^k_i(s)$ for all $k\in C$ and $i=0\ldots h-1$.\footnote{The constraints on the demand thresholds can be written as $\tau^k_{i+1}(s)\ge \tau^k_i(s)$ for all $k\in C$ and $\tau^k_{i+1}(s)> \tau^k_i(s)$  for some $k\in C$.} Let $\tau_0(s) = (0,\ldots,0)$. Let $\tau_1(\nullvendor) = (\infty,\ldots,\infty)$ for the null vendor $\nullvendor$.
We also assume that $s$ has $h$ prices $p_s^{(i)}$ for $i=1\ldots h$ such that $p_s^{(i+1)}< p_s^{(i)}$ for all $i=0\ldots h-1$.
Different vendors might have different values of $h$.

Given a allocation $\mu$, with corresponding demand vector $n(s)=(n^1(s),\ldots,n^c(s))$ for vendor $s$, $s$ offers the bundle of all items $C$ at a cumulative price $p_s^{(i^*)}$
where
$$
i^* = \max_{0\le i\le h}\{i: n^k(s) \ge \tau^k_i(s) , \forall k\in C\}.
$$
That is, $s$ offers the bundle of all products at the price corresponding to the largest demand threshold vector that is met component-wise.

If $s$ activates one of her discounts, then a buyer $b$ such that $\mu^k(b)=s$ for all $k\in C$ (i.e., $b$ buys all items from $s$) pays a price $p_s^{(i^*)}$ instead of $p_s^{(0)}$.

Let
$$
\Tau(\mu)=\{s\in\sellerset:\exists i>0 \text{ s.t. }n^k(s)\ge\tau_i^k(s),\forall k\in C\} \subseteq \sellerset
$$
be the set of vendors who activate a discount under the allocation $\mu$.

\paragraph*{Market prices}
An allocation $\mu$ determines market prices for each set of product types and each vendor.
Given $\mu$, for each $s\in\sellerset$ and $x\subseteq C$,
let $\mktpricex{s}{\mu}{x}$ denote the price at which $s$ offers the set of items $x$.
We have that
\begin{equation}
\mktpricex{s}{\mu}{x} =\left\lbrace
\begin{array}{ll}
p_s^{(i^*)} & x=C, s\in\Tau(\mu),\\
\sum_{k\in C}p_s^k  &\text{otherwise},\\
\end{array}
\right.
\end{equation}
where the definition of the index $i^*$ for vendor $s$ is given above, and corresponds to the largest threshold that is met.
In other words, vendor $s$ offers the items in set $x$ at the sum of their base prices unless $x$ corresponds to the bundle of all items and a discount threshold is met.

Similarly, an allocation $\mu$ determines market prices for each buyer.
Given $\mu$, for each $b\in\buyerset$,
let $\mktpriceb{b}{\mu}$ denote the price that $b$ pays.
We have that
\begin{equation}
\mktpriceb{b}{\mu} =\left\lbrace
\begin{array}{ll}
p_s^{(i^*)} & \text{if } \mu^k(b)=s~\forall k\in C, s\in\Tau(\mu),\\
\sum_{k\in C}p_{\mu^k(b)}^k &\text{otherwise},\\
\end{array}
\right.
\end{equation}
where the definition of the index $i^*$ for vendor $s$ is given above.

The market price paid by buyer $b$ can be seen as the sum of the market prices it pays to each vendor $s$.
Letting $x_b(s)=\{k\in C:\mu^k(b)=s \}$ be the set of items $b$ purchases from $s$,
$$
\mktpriceb{b}{\mu} = \sum_{s\in\sellerset} \mktpricex{s}{\mu}{x_b(s)}.
$$

Each vendor $s$ receives a total pay that is the sum of the market prices paid to it by each buyer,
$$
\sum_{b\in\buyerset} \mktpricex{s}{\mu}{x_b(s)}.
$$

\paragraph*{Utility}
Each buyer $b\in\buyerset$ has a valuation for each possible product choice $\bar s\in\sellersetc$.
The valuation $v_b$ of buyer $b\in\buyerset$ is a mapping from $\sellersetc$ to $\RR$,
such that, for $\bar s =(s_1,\ldots,s_c)\in\sellersetc$, $v_b(\bar s)$ is the valuation $b$ assigns to purchasing product $k$ from vendor $s_k$ for each $k\in C$.
For each  $b\in\buyerset$, let $v_b(\bar s)=0$ for $\bar s =(\nullvendor,\ldots,\nullvendor)$ (i.e., the choice not to buy any item has zero valuation).
Buyers express their valuations in term of the maximum price they are willing to pay for sets of products.
That is, $v_b(\bar s)$ represent the maximum price $b$ is willing to pay for product choice $\bar s$.

Given an allocation $\mu$, each $b\in\buyerset$ has a quasi-linear utility function given by
$$
u_b(\mu) = v_b(\mu(b)) - \mktpriceb{b}{\mu},
$$
where $\mktpriceb{b}{\mu}$ is the market price paid by $b$ under the allocation $\mu$.
Buyers play strategically, and each tries to maximize her utility.
%Given an allocation $\mu$, the price $\mktpriceb{b}{\mu}$ is computed as follows.
%If $\mu^k(b)=s$ for some $s\in\Tau(\mu)$ and all $k\in C$ then $b$ pays the price $p_s^{(i^*)}$ corresponding to the largest threshold that is met. Otherwise $b$ pays $\sum_{k\in C}p_s^{\mu^k(b)}$, that is, the sum of the base price for each single item.
%For each $s\in\sellerset$, let $x_b(s)=\{k\in C:\mu^k(b)=s \}$ be the set of items $b$ purchases from $s$.
%Recalling that $n(s)$ denotes the demand vector of $s$ under the allocation $\mu$,
%$$
%\mktpriceb{b}{\mu} = \sum_{s\in\sellerset} p_s(n(s),x_b(s)).
%$$

%
\paragraph*{Social welfare}
The \emph{social welfare} $SW(\mu)$ of an allocation $\mu$ is the sum of all buyers' utilities.
$$
SW(\mu) = \sum_{b\in\buyerset} u_b(\mu).
$$
We are interested in allocations that maximize the cumulative utility of buyers.
\begin{definition}
Allocation $\mu$ is social welfare maximizing (SWM) if $SW(\mu)\ge SW(\mu')$ for every allocation $\mu'$.
\end{definition}
In general, given a SWM allocation and the resulting market prices, one or more buyers might prefer to deviate from their product choice by strictly increasing their utility.
As our goal is to show that SMW allocations can be sustained, we turn our attention to pricing schemes in which buyers can pay prices that are higher or lower than market prices.

As a remark, social welfare maximization depends on both allocation and base prices.
In this work, we consider fixed base prices and we drop the dependency of social welfare maximization on the base prices.

\paragraph*{Implemented price}
Given her product choice, a buyer might be willing to pay a price higher than the market price, as long as the difference between her valuation and the price paid is nonnegative (and larger than that corresponding to alternative choices).
To illustrate this, consider a scenario in which the best option for buyer $b$ is to buy all items from vendor $s$ at a discounted price.
% is the product choice $\bar s =(s_1,\ldots,s_c)$ (that is, item $k$ from vendor $s_k$ for all $k\in C$) at a suitably low price.
In order to purchase the desired products at a low price, $b$ might be willing to bear some of the cost incurred by other buyers purchasing one or multiple items from $s$, which would otherwise choose other vendors.

Consider an allocation $\mu$ with market prices for buyers $\mktpriceb{b}{\mu}$.
For each $b\in\buyerset$, let $p_b$ be the price paid by buyer $b$.
$p_b$ might differ from the market price $\mktpriceb{b}{\mu}$.
We let
$$
p_b = \mktpriceb{b}{\mu} + \Delta p_b,
$$
where $\Delta p_b$ is $b$'s price difference with respect to $p_b$.
Let $p$ denote the vector of prices paid by all buyers.
We refer to the pair $(\mu,p)$ as an allocation-price pair.
The utility of $b$ under $(\mu,p)$ is given by
$$
u_b(\mu,p) = v_b(\mu(b)) - p_b = v_b(\mu(b)) - \left(\mktpriceb{b}{\mu} + \Delta p_b\right).
$$
 Let $\Delta p$ denote the vector of price differences for all buyers.
Given an allocation $\mu$, we ask whether there exist prices $p$ such that $(\mu,p)$ is stable, according to a suitable notion of stability.
We restrict our attention to price vectors such that $\sum_{b\in\buyerset} p_b = \sum_{b\in\buyerset} \mktpriceb{b}{\mu}$. %, motivated by the fact that vendors need to receive the same amount under $\mu$ and market prices and under $(\mu,p)$.
As a remark, prices $p$ are not equivalent to buyers becoming intermediaries.
In fact, a buyer might in general pay a price difference $\Delta p_b$ that is a fraction of the amount needed by another buyer, and a buyer might benefit from $\Delta p_{b'}$ paid by multiple other buyers $b'$.

\paragraph*{Stability}
%
%The strongest notion of stability for a market configuration is to exhibit the \emph{core} property~\cite{osborne}.
%A price-transfer pair $(\mu,p)$, has the core property if no coalition of buyers can increase their total utility by deviating from $\mu$.
%Maximizing the social welfare is necessary condition for the core property (otherwise all buyers can increase their social welfare by deviating to a SWM allocation).
%However, the core of a market $\market$ (the set of allocation-price pairs with the core property) can be empty (refer to the example in Section~\ref{app:example}).
We consider a notion of stability which looks at deviations by single buyers.
As a remark, stability considers an allocation-price pair $(\mu,p)$ rather than an allocation only.
The proposed notion of stability is suitable for a setting in which prices are determined by buyers' choices and buyers cannot enjoy discounts after deviation.
This assumption is in line with common group buying platforms, where buyers are given limited time to sign up to a discount offer.
Given an allocation-price pair $(\mu,p)$, there are two ways a buyer $b$ can deviate from it.
First, $b$ might deviate by changing her product choice (resulting in allocation $\mu'$ such that $\mu'(b)\neq\mu(b)$ and $\mu'(b')=\mu(b')$ for each $b'\neq b$).
In this case $b$'s utility would be given by the difference between her valuation of the newly chosen product set and the price paid.
We assume that, after deviation, $b$ has $\Delta p_b=0$ (as $\Delta p_b\neq 0$ would not constitute an unilateral action by $b$) and cannot enjoy any discount (as other buyers might not allow $b$ to enjoy discounts without paying  $\Delta p_b>0$).
Therefore, we assume that after deviation, $b$ pays the base price of the chosen products.
Second, $b$ might deviate by refusing to pay the price difference $\Delta p_b$, in full or in part.
A buyer $b$ who enjoys a discount from $s\in\Tau(\mu)$ can benefit from other buyers purchasing from vendor $s$ as they can trigger a lower price for $b$.
In this case, $b$'s payoff after deviation assumes that buyers loose incentive to buy from vendor in $s$, resulting in a price increase.
That is, we assume that a reduction of $\Delta p_b$ by $b$ results in the deviation by both subsidized  ($\Delta p_b<0$) and non-subsidized  ($\Delta p_b\ge 0$) buyers purchasing from $s$.
This assumption is motivated by the facts that buyers are unaware of each other's valuations and therefore of the total amount of subsidy needed to sustain a discount, or that a buyer refusing to pay subsidy might be banned from enjoying a discount.
In addition, it does not affect the validity of our results.
In fact, without loss of generality, we can restrict our attention to pairs $(\mu,p)$ such that no buyer $b$ receives more then the subsidy needed (i.e., $\vert\sigma_b(\mu)\vert$), and if a buyer $b'$ drops (part of) its subsidy then the current allocation cannot be sustained.
%{\red This assumption does not affect the validity of our results, as we look at the stability of pairs $(\mu,p)$ such that $\mu$ is a SWM allocation: any SWM allocation minimizes the total amount that needs to be subsidized
%in order to trigger a given price, and the deviation of each buyer providing subsidy (by paying a premium over the discounted price)  would result in a price increase.

Letting $\mu$ and $\mu'$ be respectively the allocation before and after defection by $b$, we have that $u_b(\mu') = v_b(\mu'(b)) -  \sum_{k\in C} p_{\mu'^k(b)}^k$, in both cases of $\mu'(b)\neq\mu(b)$ and $\mu'(b)=\mu(b)$.
That is, the buyer who deviates pays the base prices for the products she chooses to purchase.
The following definition formalizes the notion of stability just presented.

\begin{definition}
\label{def:stability}
An allocation-price pair $(\mu,p)$ is stable if no buyer can unilaterally and profitably deviate from it.
That is, for all $b\in\buyerset$, $u_b(\mu,p) \ge u_b(\mu')$ for each $\mu'$ such that $\mu'(b')=\mu(b')$ for each $b'\neq b$.
\end{definition}

Given allocation $\mu$, let $u^*_b(\mu)$ be the maximum utility $b$ can achieve by deviating from $\mu$, and let $\sigma_b(\mu)=u_b(\mu)-u^*_b(\mu)$ be the \emph{surplus} of $b$ under $\mu$.
If $\sigma_b(\mu)<0$ then $b$ needs to receive a subsidy ($\Delta p_b<0$) in order not to deviate from $\mu(b)$ to her best alternative.
If $\sigma_b(\mu)>0$ then $b$ might be willing to pay a subsidy to induce certain buyers not to deviate from $\mu$.
\begin{definition}
Given market $\market$ and allocation $\mu$, a price vector $p$ is stabilizing if the allocation-price pair $(\mu,p)$ is stable.
\end{definition}
Given a SWM allocation $\mu$, the existence of stabilizing prices is trivial to prove.
\begin{observation}
\label{prop:existence_stab}
For any market $\market$ and any SWM allocation $\mu$, there exist stabilizing prices $p$.
\end{observation}
We prove Observation~\ref{prop:existence_stab} by contradiction. Let $x=\sum_{b:\sigma_b(\mu)>0}\sigma_b(\mu)$ be the total subsidy available under $\mu$, and let $y=\sum_{b:\sigma_b(\mu)<0}-\sigma_b(\mu)$ be the total subsidy needed. Assume there are no stabilizing prices, that is, $x<y$.
An allocation in which each $b$ such that $\sigma_b(\mu)<0$ switches to her best alternative has social welfare at least $SW(\mu)+y-x>SW(\mu)$, generating a contradiction.
Observe that, maximizing the social welfare is sufficient but not necessary for the existence of stabilizing prices (see counterexample in Section~\ref{app:example2}).

\paragraph*{Rational and fair prices}
We are not interested in arbitrary prices, as they could be undesirable for certain buyers.
Observe that not all buyers are willing to pay $\Delta p_b>0$.
Under allocation $\mu$, a buyer $b$ is willing to pay $\Delta p_b>0$ to other buyers only if the price paid by $b$ under $\mu$ is strictly smaller than the sum of the base prices of the chosen items (i.e., $b$ buys all products from a single $s\in\Tau(\mu)$) and $b$ has positive surplus ($\sigma_b(\mu)>0$).
Moreover, a buyer is willing to pay $\Delta p_b>0$ only to benefit other buyers $b'$ that need a subsidy ($\sigma_b'(\mu)<0$) and purchase at least one product from $s$, therefore contributing to its demand vector.
In particular, $b$ is willing to pay a price difference $\Delta p_b\le \sigma_b(\mu)$ to benefit all such buyers $b'$.
If there is no buyer $b'$ such that $s\in\mu(b')$ and $\sigma_{b'}(\mu)<0$, then $\Delta p_b=0$.
In light of this, the notion of stability of Definition~\ref{def:stability} is not enough to guarantee that a stable allocation-price pair $(\mu,p)$ is desirable to buyers. Therefore, we look at price vectors $p$ that are \emph{rational} in the sense that buyers pay premiums with respect to a discounted price only to subsidize buyers that contribute to the activation of the discount.

In addition, if two buyers purchase the same items from the same vendors and have the same surplus, it would be undesirable for one of them to pay a higher price difference than the other, a principle that is encoded in our definition of \emph{fairness}.
We consider the following definitions of rationality and fairness.
\begin{definition}
\label{def:rational_price}
Given an allocation $\mu$, a price vector $p$ is rational if, for each buyer $b$,
$\Delta p_b>0$
only if 
$\sigma_b(\mu)>0$ and
$\mu^k(b)=s$ for all $k\in C$ for some $s\in\Tau(\mu)$,
 and there exists $b'$ such that $\sigma_{b'}(\mu)<0$ and $s\in \mu(b')$.
\end{definition}

\begin{definition}
\label{def:fair_price}
Given an allocation $\mu$, a rational price vector $p$ is fair if for each $b,b'\in\buyerset$ such that  $\mu(b)=\mu(b')$, $\sigma_b(\mu)>0$ and $\sigma_{b'}(\mu)>0$, the price differences $\Delta p_b$ and $\Delta p_{b'}$ paid by $b$ and $b'$ are proportional to $\sigma_b(\mu)$ and $\sigma_b'(\mu)$.
\end{definition}

Our main result (Section~\ref{market:sec:existence}) states that, given a SWM allocation $\mu$, there exists price vector $p$ that is rational according to Definition~\ref{def:rational_price} and such that the pair $(\mu,p)$ is stable according to Definition~\ref{def:stability}.
In Section~\ref{sec:computation_transf}, we focus on the computation of rational price vectors that are fair according to Definition~\ref{def:fair_price}.

\paragraph*{Transferable utility}
The price differences $\Delta p_b$ can be interpreted in terms of transferable utility.
That is, we assume that utility is transferred between buyers.
Let $\tra{b}{b'}$ denote the transfer from $b\in\buyerset$ to $b'\in\buyerset$.
Let $t$ denote the vector of transfers between all pairs of buyers.
The following definition of consistency between price and transfers defines a one-to-many correspondence between price vectors and transfer vectors.
That is, $t$ uniquely defines $\Delta p$ and therefore $p$.

\begin{definition}
\label{def:p-consistency}
Given allocation $\mu$ and price vector $p$, a transfer vector $t$ is $p$-consistent if and only if
$$
\Delta p_b =\sum_{b\in\buyerset} \tra{b}{b'},
$$
for each $b\in\buyerset$.
\end{definition}

\begin{proposition}
\label{prop:p-consistency}
Given allocation $\mu$ and price vector $p$, there exist $p$-consistent transfer vector.
Given allocation $\mu$ and transfer vector $t$, there is unique price vector $p$ such that $t$ is $p$-consistent.
\end{proposition}
The proof is given in Section~\ref{app:p-consistency}.
Therefore, $\Delta p_b$ can be written as
$$
\Delta p_b = \sum_{b'\in\buyerset}\tra{b}{b'}
$$
for a suitable $p$-consistent transfer vector $t$.
The utility of $b$ under $(\mu,p)$ can be written as
$$
u_b(\mu,p) = v_b(\mu(b)) - \left(\mktpriceb{b}{\mu} + \sum_{b'\in\buyerset}\tra{b}{b'}\right ) 
= u_b(\mu) - \sum_{b'\in\buyerset}\tra{b}{b'}.
$$

We say that a transfer vector is stabilizing if the corresponding price vector is stabilizing.
\begin{definition}
\label{def:stable_transfers}
Given market $\market$ and allocation $\mu$,
let $t$ be a transfer vector and $p$ be the unique price vector such that $t$ is $p$-consistent.
$t$ is a stabilizing transfer vector if $p$ is a stabilizing price vector.
\end{definition}

The following proposition guarantees that finding stabilizing transfers is sufficient to find stabilizing prices, provided that prices are computed according to the expression given in Definition~\ref{prop:p-consistency}.
The proof follows from Proposition~\ref{prop:p-consistency} and Definition~\ref{def:stable_transfers} and is therefore omitted.
\begin{proposition}
\label{prop:stable}
Let $t$ be a $p$-consistent transfer vector.
If $t$ is a stabilizing transfer vector then $p$ is a stabilizing price vector.
\end{proposition}

The notions of rationality and fairness introduced above can be extended to transfers.
Consider buyers $b$ and $b'$ such that $b$ buys all products from a single vendor $s\in\Tau(\mu)$, $\sigma_b(\mu)>0$ and $\sigma_{b'}(\mu)<0$.
If $s\notin\mu(b')$ then $\tra{b}{b'}=0$ as $b'$ does not affect the price $\mktpriceb{b}{\mu}$.
If $s\in\mu(b')$ then $\tra{b}{b'}$ can be positive.
In particular, $b$ is willing to pay a cumulative transfer of at most $\sigma_b(\mu)$ to all such buyers $b'$.
The reason is that these buyers might be necessary to trigger the discount $b$ currently benefits of, and they might deviate if they do not receive any transfer.

\begin{definition}
\label{def:rational_transfers}
A transfer vector $t$ is rational if, for all $b,b'\in\buyerset$,
$\tra{b}{b'}>0$
only if $\sigma_b(\mu)>0$, $\sigma_{b'}(\mu)<0$, $\mu^k(b)=s$ for all $k\in C$ and $s\in \mu(b')$ for some $s\in\Tau(\mu)$.
\end{definition}

\begin{definition}
\label{def:fair}
A rational transfer vector $t$ is fair if for each $b,b'\in\buyerset$ such that  $\mu(b)=\mu(b')$, $\sigma_b(\mu)>0$ and $\sigma_{b'}(\mu)>0$, the transfers paid by $b$ and $b'$ are proportional to $\sigma_b(\mu)$ and $\sigma_b'(\mu)$.
\end{definition}

The next two propositions guarantee that rationality and fairness of a $p$-consistent transfer vector imply rationality and fairness of $p$.
The proofs follow from Proposition~\ref{prop:p-consistency} and Definition~\ref{def:rational_transfers} and~\ref{def:fair}, and are therefore omitted.
Therefore, finding rational and fair transfer is sufficient to find rational and fair prices, provided that prices are computed according to the expression given in Definition~\ref{prop:p-consistency}.

\begin{proposition}
\label{prop:rational}
Let $t$ be a $p$-consistent transfer vector.
If $t$ is a rational transfer vector then $p$ is a rational price vector.
\end{proposition}

\begin{proposition}
\label{prop:fair}
Let $t$ be a $p$-consistent transfer vector.
If $t$ is a rational and fair transfer vector then $p$ is a rational and fair price vector.
\end{proposition}

\section{Existence of rational and stabilizing prices}
\label{market:sec:existence}

Our main result states that, maximizing the social welfare is sufficient condition for the existence of rational and stabilizing transfers.
\begin{theorem}
\label{thm:existence}
For any market $\market$ and SWM allocation $\mu$, there exist rational and stabilizing transfers.
\end{theorem}

Proposition~\ref{prop:p-consistency}, Proposition~\ref{prop:stable} and Proposition~\ref{prop:rational} imply the following corollary, which guarantees that prices can be computed according to the expression given in Definition~\ref{prop:p-consistency}.
\begin{corollary}
\label{cor:existence}
For any market $\market$ and SWM allocation $\mu$, there exist rational and stabilizing prices.
\end{corollary}

For each $s\in\Tau(\mu)$, let
$$
\posi{s}=\left\lbrace b\in\buyerset:\mu^k(b)=s~~\forall k\in C,\sigma_b(\mu)>0 \right\rbrace
$$
be the set of buyers who purchase all items from vendor $s$ (at a discounted price) and have positive surplus. For $s\notin\Tau(\mu)$ let $\posi{s}=\emptyset$.
Each $b\in\posi{s}$ is willing to pay transfers up to $\sigma_b(\mu)$ to buyers who have negative surplus and purchase at least a product $k\in C$ from $s$, for a total of
$$
\tposi{s}=\sum_{b\in\posi{s}}\sigma_b(\mu).
$$
For $s\notin\Tau(\mu)$, let $\tposi{s}=0$.

For each subset of vendors $x\subseteq\sellerset$, let
$$
\negaset{x}=\left\lbrace b\in\buyerset:\mu^k(b)\in x\}~~\forall k\in C,\sigma_b(\mu)<0 \right\rbrace
$$
be the set of buyers who purchase items from all and only the vendors in $x$ and have negative surplus.
Observe that $\negaset{x}=\emptyset$ for all $\vert x\vert>c$, so we will implicitly assume $\vert x\vert\le c$.
In order not to deviate from $\mu$ by switching to her best alternative, each $b\in\negaset{x}$ must receive a transfer of $-\sigma_b(\mu)$, for a total of
$$
\tnegaset{x}=-\sum_{b\in\negaset{x}}\sigma_b(\mu).
$$

According to Definition~\ref{def:rational_transfers}, given rational transfers $t$, if $b\in\posi{s}$ and $b'\in\negaset{x}$ for some $x\subseteq\sellerset$ such that $s\notin x$ then $\tra{b}{b'}=0$.

As a remark, given a SWM allocation $\mu$, if $s\notin\Tau(\mu)$ for all $s\in x\subseteq\sellerset$ then $\negaset{x}=\emptyset$, otherwise, allocation with higher social welfare is obtained if buyers $\negaset{x}$ switch to their best alternatives.\footnote{In particular, if $s\notin\Tau(\mu)$, then $\negaset{\{s\}}=\emptyset$. Buyers $\negaset{\{s\}}$ are the ones who purchase all items from $s$ and have negative surplus. When we restrict our attention to rational transfers, buyers $\negaset{\{s\}}$ can only receive transfer from buyers $\posi{s}$.}

\paragraph*{Group transfers}
In the proof of Theorem~\ref{thm:existence}, we will consider transfers between groups of buyers rather than transfers between single buyers.
This is enough as transfers between single buyers can be computed from group transfers in arbitrary ways (we provide a computationally efficient way which also guarantees fairness in Section~\ref{sec:computation_transf_buyers}).
In particular, for $s\in\sellerset$ and $x\subseteq\sellerset$, let
$$
\gtraset{s}{x} = \sum_{b\in\posi{s}} \sum_{b'\in\negaset{x}} \tra{b}{b'}
$$
be the total transfer from buyers $\posi{s}$ to buyers $\negaset{x}$.
If transfers $t$ are rational then $\gtraset{s}{x}=0$ whenever $s\notin x$ (and the group transfers are said to be rational).
To prove Theorem~\ref{thm:existence}, we need to show that there exist group transfers $\bar t$ such that
\begin{equation}
\label{eq:condition}
\left\lbrace
\begin{array}{ll}
\tposi{s} \ge \sum_{x:s\in x}\gtraset{s}{x} 	&\forall s\in\sellerset \\
\tnegaset{x} = \sum_{s\in x}\gtraset{s}{x} 	&\forall x\subseteq\sellerset\\
\gtraset{s}{x} = 0							&s\notin x.
\end{array}
\right.
\end{equation}
The first two constraints require that the allocation $\mu$ can be stabilized by group transfers $\bar t$, while the third constraint requires $\bar t$ to be rational.
Group transfers $\bar t$ satisfying~\eqref{eq:condition} are said \emph{rational and stabilizing}.
We consider the following definition of \emph{cross-transfer}.
\begin{definition}
For $s\in\sellerset$ and $x\subseteq\sellerset$, group transfers $\gtraset{s}{x}$ is a cross-transfer if $s\notin x$.
\end{definition}
Group transfer $\bar t$ are rational if all cross-transfers are zero.
Transfers $t$ (between buyers) are rational if and only if all cross-transfers (between groups) are zero.

Group transfers $\bar t$ and $\bar t'$ are \emph{equivalent} if buyers $\posi{s}$ pay the same transfer and buyers $\negaset{x}$ receive the same transfer under $\bar t$ and $\bar t'$.
\begin{definition}
Group transfers $\bar t$ and $\bar t'$ are equivalent if
\begin{equation*}
\left\lbrace
\begin{array}{ll}
\sum_{x\subseteq\sellerset}\gtraset{s}{x} = \sum_{x\subseteq\sellerset}\gtraset{s}{x}'\quad	&\forall s\in\sellerset \\
\sum_{s\in\sellerset}\gtra{s}{x} = \sum_{s\in\sellerset}\gtraset{s}{x}'	& \forall x\subseteq\sellerset.\\
\end{array}
\right.
\end{equation*}
\end{definition}

\paragraph*{Proof of Theorem~\ref{thm:existence}}
We assume $\mu$ is a SWM allocation.
We proceed by contradiction, making the following assumption.
\begin{assumption}
\label{ass:contr}
There are no stabilizing and rational group transfers $\bar t$.
That is, for any stabilizing group transfer $\bar t$, there are no equivalent and rational group transfers $\bar t'$.
\end{assumption}
Given a SWM allocation $\mu$ and stabilizing group transfers $\bar t$, we construct a graph $G(\bar t)$ (called the \emph{cross-transfer graph}) which encodes all cross-transfers in $\bar t$ and has no edges if and only if group transfers $\bar t$ are rational.
We then show that, given group transfers $\bar t$, there exist equivalent group transfers $\bar t'$ such that the corresponding graph $G(\bar t')$ is directed and acyclic.
Assumption~\ref{ass:contr} implies that any such $G(\bar t')$ has edges,
and we complete the proof by showing that allocation $\mu'$ with $SW(\mu')>SW(\mu)$ can be obtained, generating a contradiction with the assumption that $\mu$ is a SWM allocation.

\begin{definition}
\label{def:ctgraph}
Given group transfers $\bar t$, the cross-transfer graph $G(\bar t)$ is the directed graph with node set equal to $\sellerset$, and directed edge $(s,s')$ if and only if
there exist $x\subseteq S$ such that $s\notin x, s'\in x, \gtraset{s}{x}>0$.
\end{definition}

\begin{figure}
\centering
\includegraphics[scale=0.5]{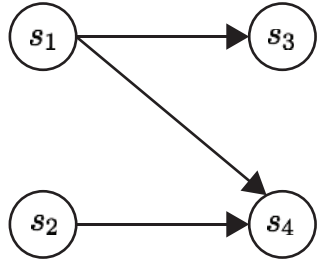}
\caption{{
Example of a cross-transfer graph. Assume that $\sellerset=\{s_1,s_2,s_3,s_4\}$, and that the only nonzero cross-transfers are $\gtraset{s_1}{x'}>0$ for $x'=\{s_3,s_4\}$ and $\gtraset{s_2}{x''}>0$ for $x''=\{s_4\}$. According to Definition~\ref{def:ctgraph}, $G(\bar t)$ has nodes $\{s_1,s_2,s_3,s_4\}$ and directed edges $\{(s_1,s_4),(s_1,s_4),(s_2,s_4)\}$.
}}
\label{fig:graph1}
\end{figure}

In words, in $G(\bar t)$ there is an edge from $s\in\sellerset$ to $s'\in\sellerset$ if buyers $\posi{s}$ pay a cross-transfer to buyers $\negaset{x}$ for some $x\subseteq\sellerset$ such that $s\notin x, s'\in x$. An example of cross-transfer graph is given in Figure~\ref{fig:graph1}.

The following results state that rational group transfers correspond to cross-transfer graphs with no edges,
and that we can restrict our attention to directed acyclic graphs.

\begin{lemma}
\label{lem:empty}
Group transfers $\bar t$ are rational if and only if $G(\bar t)$ has no edge.
\end{lemma}

\begin{lemma}
\label{lem:DAG}
Given group transfers $\bar t$, there exist equivalent group transfers $\bar t'$ such that the corresponding cross-transfer graph $G(\bar t')$ is acyclic.
\end{lemma}
The proof of Lemma~\ref{lem:empty} follows by the definition of cross-transfer graph and is therefore omitted.
The proof of Lemma~\ref{lem:DAG} is given in Section~\ref{app_lem:DAG}.

Without loss of generality, consider stabilizing group transfers $\bar t$ and assume that $G(\bar t)$ is a directed acyclic graph.
By Assumption~\ref{ass:contr}, there are no equivalent group transfers $\bar t'$ such that $G(\bar t')$ has no edge.
A vendor $s\in\sellerset$ is called a \emph{source} node if there is no edge $(s',s)$ in $G(\bar t')$, and an \emph{internal} node otherwise.
Let $\sellerset^{SRC}\subseteq\sellerset$ be the set of vendors corresponding to source nodes in $G(\bar t')$.
Let $\sellerset^{IN}\subseteq\sellerset$ be the set of vendors corresponding to internal nodes in $G(\bar t')$.
Let
$$
\mathcal{N}^{IN} = \bigcup \{\negaset{x} \text{ s.t. } x\subseteq\sellerset^{IN} \} 
$$
be the set of buyers who purchase products only from vendors $\sellerset^{IN}$ (and possibly some product from the null vendor $\nullvendor$) and have negative surplus.
By Lemma~\ref{lem:DAG}, we can assume without loss of generality that all buyers who receive transfer purchase products only from vendors $\sellerset^{IN}$.
Let
$$
\mathcal{P}^{IN} = \bigcup \{\posi{s} \text{ s.t. } s\in\sellerset^{IN} \}
$$
be the set of buyers who buy all items $C$ from a single vendor in $\sellerset^{IN}$
and have positive surplus.
Similarly, let
$$
\mathcal{P}^{SRC} = \bigcup \{\posi{s} \text{ s.t. } s\in\sellerset^{SRC} \}.
$$
According to $G(\bar t')$, buyers $\mathcal{P}^{IN}$ are not able to pay the total amount of transfer needed by buyers $\mathcal{N}^{IN}$, and additional transfer from $\mathcal{P}^{SRC}$ is needed (observe that the latter buyers get no benefit from the product choice of buyers $\mathcal{N}^{IN}$).
Under Assumption~\ref{ass:contr}, letting
$$
X = \sum_{b\in\mathcal{P}^{IN}} \sigma_b(\mu)
\qquad
\text{and}
\qquad
Y = -\sum_{b\in\mathcal{N}^{IN}} \sigma_b(\mu),
$$
be the amounts of transfer made available by $\mathcal{P}^{IN}$ and needed by $\mathcal{N}^{IN}$ respectively, we have that $X<Y$.
Consider the allocation $\mu'$ in which \emph{all} buyers $\mathcal{N}^{IN}$ and $\mathcal{P}^{IN}$ deviate to their best alternatives\footnote{Allocation $\mu'$ is the results of a deviation from $\mu$ by multiple buyers. We do not directly use this deviation to proof the stability of allocation-transfer pair (whose definition looks at unilateral deviations). We use $\mu'$ to derive a contradiction on the assumption that $\mu$ is a SWM allocation.}.
Buyers $\mathcal{N}^{IN}$ incur a cumulative gain of at least $Y$  (the gain would be strictly greater than $Y$ if some new threshold is activated for these buyers, after deviation\footnote{Even if for the sake of stability buyers cannot enjoy discounts after deviation, here we consider that discount thresholds might be triggered as we are interested in computing the social welfare of $\mu'$.}).
Buyers $\mathcal{P}^{IN}$ can either gain or loose utility after deviation, but each cannot incur a loss larger than $\sigma_b(\mu)$, resulting in an upper bound of $X$ on the cumulative loss.\footnote{It is necessary to assume that also buyers $\mathcal{P}^{IN}$ deviate to their best alternatives, as their surplus $\sigma_b(\mu)$ depends on their best alternatives given the allocation-transfer pair $(\mu,t)$.}
Buyers $\mathcal{P}^{SRC}$ cannot loose utility, as no buyer deviates from sellers $\sellerset^{SRC}$ as we consider deviations by buyers $\mathcal{N}^{IN}$.
All remaining buyers are the ones in $\negaset{x}$ for $x\subseteq\sellerset$ such that $x\cap \sellerset^{SRC}\neq\emptyset$ (i.e., buyers with negative surplus who do not buy all items from $\sellerset^{IN}$) and all buyers with nonnegative surplus who are not enjoying any discount.
Since these buyers do not enjoy the discounts by vendors $\sellerset^{IN}$, they cannot loose utility in $\mu'$ with respect to $\mu$.
We have that $SW(\mu')\ge SW(\mu) + Y - X > SW(\mu)$, generating a contradiction with the assumption that $\mu$ is a SWM allocation.

\section{Computation of rational, fair and stabilizing prices}
\label{sec:computation_transf}
Given a market $\market$ and a SWM allocation $\mu$, Theorem~\ref{thm:existence} guarantees the existence of rational and stabilizing transfers, and Corollary~\ref{cor:existence} guarantees the existence of corresponding buyers' prices.
In this section we present an efficient procedure to compute rational and stabilizing transfers that are also fair according to Definition~\ref{def:fair}.
Then, Proposition~\ref{prop:p-consistency} and Proposition~\ref{prop:fair} guarantee that buyers' prices that are fair according to Definition~\ref{def:fair_price} can be computed via the expression given in Definition~\ref{def:p-consistency},
$$
\Delta p_b =\sum_{b\in\buyerset} \tra{b}{b'}.
$$

%
%For ease of notation, let $\sellerset=\{0,1,\ldots,M\}$, where $s=0$ represents the ``null vendor'' (corresponding to the choice not to purchase a product).
Recall that $N$, $M$ and $c$ are the numbers of buyers, sellers and product types, respectively.
We assume that a SWM allocation $\mu$ is given, and we proceed as follows.
In Section~\ref{sec:computation_transf_group} we show how to compute rational and stabilizing group transfers $\gtraset{s}{x}$ from $\posi{s}$ to $\negaset{x}$ for all $s\in\sellerset$, $x\subseteq\sellerset$ ($\vert x\vert\le c$), via the max-flow Ford-Fulkerson algorithm on a flow network such that feasible flows are in one-to-one correspondence with rational group transfers.
Let $T=\sum_{x}\tnegaset{x}$ be the total transfer needed by buyers with negative surplus (that is, all $b$ such that $\sigma_b(\mu)<0$) and who are not purchasing both items from the same vendor. Assuming that prices and valuations are constant in $N$ and $M$, and observing that $\vert\cup_{x}\negaset{x}\vert \le N$, we have that $T=\bigO(N)$,
and rational and stabilizing group transfers can be computed in time $\bigO(TM^c)=\bigO(NM^c)$.

Given rational and stabilizing group transfers, in Section~\ref{sec:computation_transf_buyers} we show how to compute rational and stabilizing transfers, and therefore buyers' prices that are fair according to Definition~\ref{def:fair}.
This requires time $\bigO(N^2+NM^{c-1})$, for an overall time $\bigO(N^2+NM^c)$.

\subsection{Step 1: rational and stabilizing group transfers}
\label{sec:computation_transf_group}
\begin{figure}
\centering
\includegraphics[scale=0.5 ]{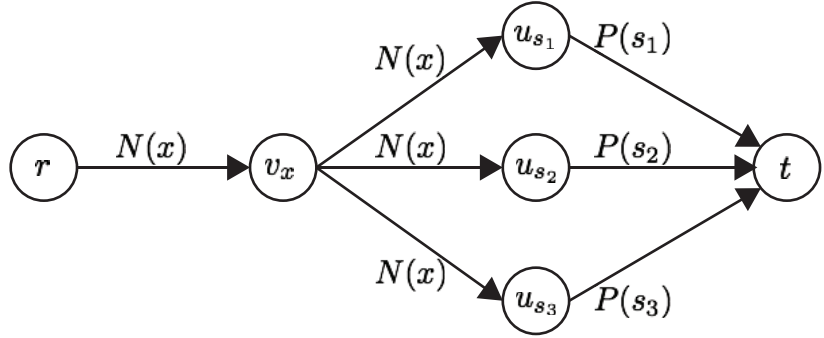}
\caption{{\small Scheme of the flow network $\network$.
A single node $v_{x}$ for a set $x=\{s_1,s_2,s_3\}\subseteq\sellerset$ is represented.
There is an edge from the source $r$ to $v_{x}$ with capacity $\tnegaset{x}$, to accommodate the total transfer needed by $\negaset{x}$.
For $i=1,2,3$, there is an edge from $v_{x}$  to $u_{s_i}$ with capacity $\tnegaset{x}$, to accommodate the transfer from $\posi{s_i}$ to $\negaset{x}$.
For $i=1,2,3$, there is an edge from $u_{s_i}$ to the sink $t$ with capacity $\tposi{s_i}$, to accommodate the total transfer from $\posi{s_i}$ (transfer not only to $\tnegaset{x}$).
}}
\label{fig:flow1}
\end{figure}

We consider the following flow network $\network$ (refer to Figure~\ref{fig:flow1}).
Nodes are as follows.\\
-- A single source node $r$, and a single sink node $t$.\\
-- A node $v_{x}$ for each $x\subseteq\sellerset$, $\vert x\vert\le c$, corresponding to $\negaset{x}$. There are $\bigO(M^c)$ such nodes.\\
-- A node $u_s$ for each $s\in\sellerset$, corresponding $\posi{s}$. There are $M$ such nodes.\\
%-- A single source node $t$.\\
Edges and capacities are as follows.\\
-- For each node $v_{x}$, an edge from $r$ to $v_{x}$ with capacity $\tnegaset{x}$. Flow from $s$ to $v_{x}$ represents the total transfer to $\negaset{x}$. There are $\bigO(M^c)$ such edges.\\
-- For each node $v_{x}$, and edge from $v_{x}$ to $u_s$ for all $s\in x$, each with capacity $\tnegaset{x}$. Flow from $v_{x}$ to $u_s$ represents the group transfer from $\posi{s}$ to $\negaset{x}$. There are $\bigO(M^c)$ such edges (as each node $v_x$ has at most a constant number $c$ of outgoing edges).\\
-- For each node $u_s$, an edge from $u_s$ to $t$ with capacity $\tposi{s}$. Flow from $u_s$ to $t$ represents the total transfer given by $\posi{s}$. There are $M$ such edges.

Given a flow $f$ on the network $\network$, $f(x,y)$ represents the flow from node $x$ to node $y$.
Let $\fflows$ be the set of all feasible flows on $\network$ and $\rtrans$ be the set of all rational group transfers in the market $\market$ (given the SWM allocation $\mu$). Consider the mapping $\omega:\fflows \to \rtrans$ such that a feasible flow $f\in\fflows$ is mapped to group transfers $\bar t=\omega(f)$ such that:
\begin{equation*}
\left\lbrace
\begin{array}{ll}
\gtraset{s}{x}=f(v_{x},u_s) 	&x\subseteq\sellerset,\vert x\vert\le c, s\in\sellerset \text{ such that there is edge }(v_{x},u_s)\text{ in }\network\\
\gtraset{s}{x} = 0							&\text{otherwise}.
\end{array}
\right.
\end{equation*}
Observe that the capacity constraints on edges $(u_s,t),s\in\sellerset$ imply that $\sum_{x}\gtraset{s}{x}\le \tposi{s}$ for all $s\in\sellerset$. $\bar t$ is rational as in $\network$ there is no edge $(v_{x},u_s)$ for $s\notin x$.

%\begin{proposition}
%\label{prop:flow1}
%The mapping $\omega:\fflows \to \rtrans$ is a bijection.
%\end{proposition}
%The proof is straightforward and therefore it is omitted.
%The following results states that maximum flows correspond to stabilizing group transfers.
\begin{proposition}
\label{prop:flow2}
The mapping $\omega:\fflows \to \rtrans$ is a bijection.
Let $f^*$ be a maximum flow on $\network$. Then, $\omega(f^*)$ defines rational and stabilizing group transfers.
\end{proposition}
%\begin{proof}
%It is straightforward to see that $\omega$ is a bijection.
%Let $\bar t=\omega(f^*)$. $\bar t$ are rational group transfers (as  $\omega$ is a bijection). Suppose by contradiction that $\bar t$ is not stabilizing, that is, condition~\eqref{eq:condition} does not hold for $\bar t$. Recall that condition~\eqref{eq:condition} read as
%\begin{equation*}
%\left\lbrace
%\begin{array}{ll}
%\tposi{j} \ge \sum_{h\neq j}\gtra{j}{j}{h} 	&\forall j\in\sellerset \\
%\tnega{h}{k} = \gtra{h}{h}{k} + \gtra{k}{h}{k}\quad 	&\forall h,k\in\sellerset \\
%\gtra{j}{h}{k} = 0							&j\neq h, j\neq k.
%\end{array}
%\right.
%\end{equation*}
%First, suppose that $\tposi{j} < \sum_{h\neq j}\gtra{j}{j}{h}$ for some $j\in\sellerset$. This would imply that the flow entering node $u_j$ is larger than the capacity of the edge $(u_j,t)$, generating a contradiction with the feasibility of the maximum flow $f^*$.
%Second, suppose that $\tnega{h}{k} > \gtra{h}{h}{k} + \gtra{k}{h}{k}$ for some $h\neq k$.
%This would imply that every flow $f'$ on $\network$ is smaller than $\sum_{h\neq k}\tnega{h}{k}$, and therefore there exist no group transfers $\bar t'$ such that $\tnega{h}{k} = \gtra{h}{h}{k}' + \gtra{k}{h}{k}'$ for all $h,k\in\sellerset$, generating a contradiction with Theorem~\ref{thm:existence} (as feasible flows and rational group transfers are in one-to-one correspondence).
%Rationality of $\bar t$ implies that $\gtra{j}{h}{k} = 0$ if $j\neq h,k$.
%\end{proof}
The proof is given in Section~\ref{app:propflow2}.
We can therefore compute rational and stabilizing group transfers via the Ford-Fulkerson algorithm for the maximum flow (see for example~\cite{kleinbergalgorithm}).
To bound the running time of the algorithm, we assume that the capacities of all edges in $\network$ are integer, that is, all terms $\tposi{s}$ and $\tnegaset{x}$ are integer. This is the case if valuations and prices are multiples of the same unit (e.g., dollars or cents). For a network with $n$ nodes, $e$ edges, integer capacities, and the total capacity of the edges exiting the source equal to $T$, the running time of the  algorithm is $\bigO((m+n)T)$.
In $\network$, we have that $n=\bigO(M^c)$, $e=\bigO(M^c)$, and $T=\sum_{x}\tnegaset{x}$.
Therefore, stabilizing group transfers can be computed in time $\bigO(TM^c)$.
If we assume that prices and valuations are $\bigO(1)$ (that is, constant in the market size $N$, $M$), we have that $T=\bigO(N)$ (as $\vert\cup_{x}\negaset{x}\vert\le N$) and that $\bigO(TM^c)=\bigO(NM^c)$.

\subsection{Step 2: computation of buyers' prices}
\label{sec:computation_transf_buyers}
In this section we show how rational, fair and stabilizing buyers' prices can be computed from rational and stabilizing group transfers.
In particular, we show how to compute fair transfers between buyers,
and buyers' prices uniquely follow from market prices as
$$
p_b = \mktpriceb{b}{\mu} + \Delta p_b = \mktpriceb{b}{\mu} + \sum_{b\in\buyerset} \tra{b}{b'}.
$$
Observe that each buyer $b\in\buyerset$ belongs at most to a single set $\posi{s}$ for some $s\in\sellerset$ or to a single set $\negaset{x}$ for some $x\subseteq\sellerset$, $\vert x\vert\le c$.
We consider the following definition of fairness, equivalent to Definition~\ref{def:fair} when we restrict our attention to stabilizing transfers.

\begin{definition}
Given a market $\market$ and a SWM allocation $\mu$, rational and stabilizing transfers $t$ (with corresponding group transfers $\bar t$) are fair if, for each $s\in\sellerset$ such that $\posi{s}\neq\emptyset$ and each $b\in\posi{s}$, the total transfer paid by $b$ is
$$
\sum_{b'\in\buyerset}\tra{b}{b'} = \sigma_b(\mu)\sum_{x\subseteq\sellerset}\gtraset{s}{x} / \tposi{s}.
$$
\end{definition}
Observe that all buyers $\posi{s}$ are required to pay a cumulative transfer of $\sum_{x}\gtraset{s}{x}$ to buyers $\bigcup_{x}\negaset{x}$, out of an available cumulative surplus of $\tposi{s} = \sum_{b\in\posi{s}}\sigma_b(\mu)$.
Under rational, fair and stabilizing group transfers $\bar t$,
Condition~\eqref{eq:condition} guarantees that no buyer with $\sigma_b(\mu)>0$ pays more than $\sigma_b(\mu)$, and that each buyer with $\sigma_b(\mu)<0$ can receive the required side-payment.

We now present our algorithm to compute rational and fair stabilizing transfers from rational and stabilizing group transfers.
First, $\tra{b}{b'}$ is initialized at zero for each $b,b'\in\buyerset$.
Fair transfers from buyers $\posi{s}$ (for a fixed $s\in\sellerset$ such that $\posi{s}\neq\emptyset$) are computed by algorithm $\alg_1$ (in Table~\ref{alg:first}), as follows.

Assume that $\gtraset{s}{x}>0$ for $x=x_1,\ldots,x_h$ (with $s\in x_k$ for all $k=1,\ldots,h$), as output by the algorithm in Section~\ref{sec:computation_transf_group}.
Observe that $h=\bigO(M^{c-1})$ as we are considering sets $x$ such that $\vert x\vert\le c$ and $s\in x$.

For each $b\in\posi{s}$, at any given point in the execution of the algorithm, $\tilde\sigma_b$ denotes $b$'s residual surplus, that is, the amount $b$ has still available to make side-payments.
At initialization, let $\tilde\sigma_b = \sigma_b(\mu)>0$.
Transfers to buyers $\negaset{x_k}$ are computed in phases, in increasing order of $k$.
At each phase $k=0,\ldots,h$, let $\alpha = \gtraset{s}{x_k}/\sum_{b\in\posi{s}}\tilde\sigma_b(\mu)$ be the ratio between the group transfer from $\posi{s}$ to $\negaset{x_k}$ and the residual surplus of $\posi{s}$, and let $\beta = \gtraset{s}{x_k} / \tnegaset{x_k}$ be the fraction of transfer that $\negaset{x_k}$ receives from $\posi{s}$, out of the total transfer from $\cup_{s'\in x_k}\posi{s'}$.
Algorithm $\alg_2$ in Table~\ref{alg:second} computes transfers between buyers $\posi{s}$ to buyers $\negaset{x_k}$ such that each $b\in\posi{s}$ transfers $\alpha\tilde\sigma_b(\mu)$ and each $b'\in\negaset{x_k}$ receives $-\beta\sigma_{b'}(\mu)$.
Before increasing the value of $k$, each $b\in\posi{s}$ updates her residual surplus to $(1-\alpha)\tilde\sigma_b(\mu)$.

The correctness of algorithm $\alg_2$ is straightforward.
Given this, the correctness of algorithm $\alg_1$ follows by observing that, for each $s\in\sellerset$ and $b\in\posi{s}$, $b$'s transfer in each instance of algorithm $\alg_2$ never exceed $\tilde\sigma_b$, and that for each $x\subseteq\sellerset$, $\vert x\vert\le c$ and $b'\in\negaset{x}$, $b'$ receives a total of $-\sigma_b(\mu)$ in the (at most) $c$ instances of algorithm $\alg_2$ she is involved in.

%
%\floatname{algorithm}{Table}
%\begin{algorithm}[h!]
%\caption{Algorithm $\alg_1$, transfers from buyers in $\posi{j}$}
%\label{alg:first}
%\begin{algorithmic}
%\STATE{\textbf{Input:} $\gtra{j}{j}{k}$ for all $k=0,\dots,M$, $\sigma_b(\mu)$ for all $b\in\buyerset$}
%\STATE{\textbf{Initialize:} $\tilde\sigma_b = \sigma_b(\mu)$ for each $b\in\posi{j}$}
%\FOR{$k=0,\ldots,M$}
%\IF{$\gtra{j}{j}{k}>0$}
%\STATE{$s\leftarrow\sum_{b\in\posi{j}}\tilde\sigma_b$;$\alpha \leftarrow \gtra{j}{j}{k} / s$; $\beta \leftarrow \gtra{j}{j}{k} / (\gtra{j}{j}{k}+\gtra{k}{j}{k})$}
%%\STATE{$\alpha \leftarrow \gtra{j}{j}{k} / s$}
%%\STATE{$\beta \leftarrow \gtra{j}{j}{k} / (\gtra{j}{j}{k}+\gtra{k}{j}{k})$}
%\STATE{Algorithm $\alg_2$ with input $\{\alpha\tilde\sigma_b:b\in\posi{s}\}$, $\{-\beta\tilde\sigma_{b'}:b'\in\nega{j}{k}\}$}
%\FOR{$b\in\posi{s}$}
%\STATE{$\tilde\sigma_b \leftarrow (1-\alpha)\tilde\sigma_b$}
%\ENDFOR
%\ENDIF
%\ENDFOR
%\end{algorithmic}
%\end{algorithm}
%
%

\begin{algorithm}[t]
\SetAlgoNoLine
\KwIn{$\gtra{j}{j}{k}$ for all $k=0,\dots,M$, $\sigma_b(\mu)$ for all $b\in\buyerset$.}
\textbf{Initialize:} $\tilde\sigma_b = \sigma_b(\mu)$ for each $b\in\posi{s}$\;
\For{$k=0,\ldots,M$
    }{
      \If{$\gtra{j}{j}{k}>0$\;
      }{
        $s\leftarrow\sum_{b\in\posi{s}}\tilde\sigma_b$\;
	$\alpha \leftarrow \gtra{j}{j}{k} / s$\;
	$\beta \leftarrow \gtra{j}{j}{k} / (\gtra{j}{j}{k}+\gtra{k}{j}{k})$\;
	Algorithm $\alg_2$ with input $\{\alpha\tilde\sigma_b:b\in\posi{s}\}$, $\{-\beta\tilde\sigma_{b'}:b'\in\nega{j}{k}\}$\;
	\For{$b\in\posi{s}$
    	    }{
	    $\tilde\sigma_b \leftarrow (1-\alpha)\tilde\sigma_b$\;
	}
      }
        }
\caption{Algorithm $\alg_1$, transfers from buyers in $\posi{s}$}
\label{alg:first}
\end{algorithm}

%
%\floatname{algorithm}{Table}
%\begin{algorithm}[h!]
%\caption{Algorithm $\alg_2$: transfers between buyers}
%\label{alg:second}
%\begin{algorithmic}
%\STATE{\textbf{Input:} amounts offered $\{x_1,\ldots,x_n\}$ by $\{b_{h_1},\ldots,b_{h_n}\}$; requested $\{y_1,\ldots,y_m\}$ by  $\{b_{k_1},\ldots,b_{k_m}\}$}
%\STATE{\textbf{Onput:} transfers $\tra{h_i}{k_{\ell}}$ for $i=1,\ldots,n$ and $\ell=1,\ldots,m$}
%\STATE{\textbf{Initialize:} $i=1$, $\ell=1$}
%\WHILE{$\ell\le m$ \AND $y_{\ell}>0$}
%\IF{$x_i\ge y_{\ell}$}
%\STATE{$\tra{h_i}{k_{\ell}} \leftarrow y_{\ell}$; $x_i \leftarrow x_i - y_{\ell}$; $\ell\leftarrow\ell+1$}
%%\STATE{$x_i \leftarrow x_i - y_{\ell}$}
%%\STATE{$\ell\leftarrow\ell+1$}
%\ELSE
%\STATE{$\tra{h_i}{k_{\ell}} \leftarrow x_i$; $y_{\ell} \leftarrow y_{\ell} - x_i $; $i\leftarrow i+1$}
%%\STATE{$y_{\ell} \leftarrow y_{\ell} - x_i $}
%%\STATE{$i\leftarrow i+1$}
%\ENDIF
%\ENDWHILE
%%\RETURN $t$
%\end{algorithmic}
%\end{algorithm}

\begin{algorithm}[t]
\SetAlgoNoLine
\KwIn{amounts offered $\{x_1,\ldots,x_n\}$ by $\{b_{h_1},\ldots,b_{h_n}\}$; requested $\{y_1,\ldots,y_m\}$ by  $\{b_{k_1},\ldots,b_{k_m}\}$.}
\KwOut{transfers $\tra{h_i}{k_{\ell}}$ for $i=1,\ldots,n$ and $\ell=1,\ldots,m$}
\textbf{Initialize:} $i=1$, $\ell=1$\;
\While{($\ell\le m$) \text{and} ($y_{\ell}>0$)
	}{
	\eIf{$x_i\ge y_{\ell}$
		}{
		$\tra{h_i}{k_{\ell}} \leftarrow y_{\ell}$\;
		$x_i \leftarrow x_i - y_{\ell}$\;
		$\ell\leftarrow\ell+1$\;
		}
		{%else
		$\tra{h_i}{k_{\ell}} \leftarrow x_i$\;
		$y_{\ell} \leftarrow y_{\ell} - x_i $\;
		$i\leftarrow i+1$\;
		}
	}
\caption{Algorithm $\alg_2$, transfers from buyers in $\posi{s}$ to buyers in $\nega{j}{k}$}
\label{alg:second}
\end{algorithm}

\paragraph*{Time complexity}
Let $T_{\alg_1}(s)$ and $T_{\alg_2}(s,x)$ be the number of operations required, respectively, by algorithm $\alg_1$ for buyers in $\posi{s}$,
and by algorithm $\alg_2$ to compute transfers from $\posi{s}$ to $\negaset{x}$. The total time to compute fair, rational and stabilizing transfers is $T(M,N)=\bigO(N^2) + \sum_{s\in\sellerset} T_{\alg_1}(s)$, where the first terms accounts for the initialization of $t$.

We have that $T_{\alg_2}(s,x) = \bigO(\vert\posi{s}\vert + \vert\negaset{x}\vert)$, as during each iteration of the \emph{while} loop, one of the indexes $i$ and $\ell$ increases by one, and each iteration requires a constant number of operations.

To upper bound $T_{\alg_1}(s)$, each iteration of the \emph{for} loop requires $\bigO(\vert\posi{s}\vert)$ operations to compute $s$, and $T_{\alg_2}(s,x)$ operations for the execution of algorithm $\alg_2$. Therefore, the cumulative running time is upper bounded by
\begin{align*}
T(M,N)	&=\bigO(N^2) + \sum_{s\in\sellerset} T_{\alg_1}(s) \\
		&= \bigO(N^2) + \sum_{s\in\sellerset} \sum_{\vert x\vert\le c:s\in x}\left( T_{\alg_2}(s,x)+\bigO(\vert\posi{s}\vert) \right)\\
		&= \bigO(N^2) + \sum_{s\in\sellerset} \sum_{\vert x\vert\le c:s\in x}
		 \bigO(\vert\posi{s}\vert + \vert\negaset{x}\vert) \\
		&= \bigO(N^2) + \sum_{s\in\sellerset} \bigO(M^{c-1}) \bigO(\vert\posi{s}\vert) + \sum_{s\in\sellerset} \sum_{\vert x\vert\le c:s\in x} \bigO(\vert\negaset{x}\vert) \\
		&= \bigO(N^2) + \bigO(M^{c-1}N) + \bigO(N) = \bigO(N^2+M^{c-1}N)
\end{align*}
as $\sum_{s\in\sellerset}\vert\posi{s}\vert\le N$, $\sum_{s\in\sellerset} \sum_{\vert x\vert\le c:s\in x} \vert\negaset{x}\vert\le cN$, and $\vert\{x\subseteq\sellerset:s\in x \}\vert= \bigO(M^{c-1})$.

The time to compute buyers' prices from transfers is $\bigO(N^2)$, as
$$
p_b = \mktpriceb{b}{\mu} + \sum_{b,b'\in\buyerset} \tra{b}{b'}.
$$
Combining with the result in Section~\ref{sec:computation_transf_group}, fair, rational and stabilizing buyers' prices can be computed in time $\bigO(N^2+NM^{c})$ given a SWM allocation.

\section{Computation of social welfare maximizing allocation}
\label{sec:swm}
A natural approach to compute a SWM allocation is to formulate a mixed integer program (see~\cite{rothblum}) in which, for each $b\in\buyerset$ and $\bar s\in\sellersetc$, a binary assignment variable $x_{b,\bar s}$ indicates whether $\mu(b)=\bar s$, and, for each $s\in\sellerset$, a binary variable $z_{s,i}$ indicates whether the demand of vendor $s$ meets the threshold $\tau_i(s)$ (and the corresponding discount is triggered).
These would account to $NM^c+H$ integer variables, where $H\ge M$ is the total number of discount thresholds of all vendors, and a running time exponential in this quantity.
A relaxation of the problem by letting each assignment variable lay in the interval $[0,1]$ would leave only $H$ integer variables (and the computational complexity exponential in $M$). However, the existence of an integral solution (corresponding to a valid allocation) is an open question.

Instead, we follow a different approach, similar to~\cite{luboutlier}. Conditional on the number of buyers $n(\bar s)=\vert \hat\mu(\bar s) \vert$ for each  $\bar s\in\sellersetc$ (which we refer to as a \emph{partition} of the buyers), we compute a SWM allocation via the Ford-Fulkerson algorithm for the max-flow with min-cost in time $\bigO(N^2M^c)$.
Then by considering all feasible allocations $\{n(\bar s):\bar s\in\sellersetc\}$ (that are however exponential in $M^c$), we determine the SWM allocation.

Let
$\feasible = \left\{  \{n(\bar s):\bar s\in\sellersetc\} :  \sum_{\bar s\in\sellersetc}n(\bar s)=N \right\}$
be the set of all \emph{feasible} partitions, that is, partitions such that each of the $N$ buyers can be assigned to a single set of vendors.

\begin{figure}
\centering
\includegraphics[scale=0.5]{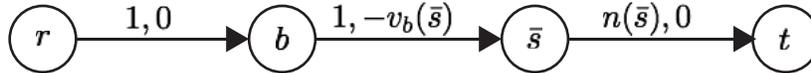}
\caption{{\small Scheme for the flow network $\network(\pi)$. Nodes for a single buyer $b\in\buyerset$ and a single set $\bar s\in\sellersetc$ are represented.
The cost of $-v_{b}(\bar s)$ of the edge from $b$ to $\bar s$ is the opposite of the valuation buyer $b$ gives to product choice $\bar s$.
}}
\label{fig:flow2}
\end{figure}

Fix $\pi\in\feasible$, and define a flow network $\network(\pi)$ as follows (see Figure~\ref{fig:flow2}).
Nodes are the following.\\
-- A single source node $r$, and a single sink node $t$.\\
-- For each $b\in\buyerset$, a node $b$. There are $N$ such nodes.\\
-- For each $\bar s\in\sellersetc$, a node $\bar s$. There are $M^c$ such nodes.\\
The edges, with corresponding capacities and costs, are as follows.\\
-- For each $b\in\buyerset$, an edge from $r$ to $b$, with capacity $1$ and cost $0$. A unit of flow on this edge represents $b$ being assigned to some product choice. There are $N$ such edges.\\
-- For each $b\in\buyerset$ and $\bar s\in\sellersetc$, an edge from $b$ to $\bar s$ with capacity $1$ and cost $-v_{b}(\bar s)$, that is, the opposite of the valuation buyer $b$ gives to product choice $\bar s$. A unit flow on this edge represents buyer $\mu(b)=\mu$. There are $NM^c$ such edges.\\
-- For each $\bar s\in\sellersetc$, an edge from $\bar s$ to $t$ with capacity $n(\bar s)$ and cost $0$. An integral flow on this edge represents the total number of buyers choosing $\bar s$. There are $(M+1)^c$ such edges.

Feasible integral flows on $\network(\pi)$ are in one-to-one correspondence with allocations conditional on $\pi$.
Let $\mu(f)$ be the allocation corresponding to flow $f$.
Given an integral flow $f$ on $\network(\pi)$, its value equals the number of buyers that are matched to vendor pairs in $\mu(f)$, and its cost equals the negative of the total valuation by buyers under $\mu(f)$.
Every max-flow $f$ on $\network(\pi)$ has value $N$, that is, each buyer is matched to a vendor pair under $\mu(f)$.
Given $\pi\in\feasible$,
%and letting $\Tau(\pi)$ be the set of vendors who activate their bundle discount under $\pi$,
the total price paid by buyers is constant for each max-flow $f$ on $\network(\pi)$.
%$$
%\sum_{\bar s\in\sellersetc} n(\bar s) (\pA{j} + \pB{k}) - \sum_{j\in\Tau(\pi)} \njk{j}{j} (\pA{j} + \pB{j} - \pAB{j}). 
%$$
Therefore, maximizing the social welfare of allocation conditional on $\pi$ corresponds to minimizing the cost of an integral max-flow on $\network(\pi)$. 
The total numbers of nodes and edges in $\network(\pi)$ are respectively $n=\bigTheta(N+M^c)$ and $e=\bigTheta(NM^c)$, and the total capacity of the edges exiting the source is $T=N$.
Since all capacities are integer, the Ford-Fulkerson algorithm finds an integral max-flow with minimum cost in time $\bigTheta(T(n+m))=\bigTheta(N^2M^c)$.

To determine the SWM allocation of $\market$, for each $\pi\in\feasible$ we need to determine, a SWM allocation conditional on $\pi$, for an overall time $\bigTheta(N^2M^c\vert\feasible\vert)$.
However, this is dominated by a term $N^{M^c}$ (see Section~\ref{app:expo}).

Getting rid of the exponential dependency in $M$ does not seem possible, due to the theoretical hardness of the problem. In fact, fixed $x>0$, deciding whether there exists allocation $\mu$ with $SW(\mu)\ge x$ is NP-hard, (by a reduction from the Knapsack problem, as noted by~\cite{luboutlier}).
Even if computationally demanding even for small $M$, the proposed solution requires time polynomial in the number of buyers $N$.
Our solution is significantly more efficient than both the exhaustive maximization of social welfare over all $M^{cN}$ allocations, and solving the integer problem above (both exponential in $N$).
Moreover, $M$ could in general be considered much smaller than $N$, or even constant.

\section{Discussion}
\label{sec:discussion}
It is an open question whether Theorem~\ref{thm:existence} holds in the case of arbitrary price schedules, where a vendor might have several discounted prices on sets of products, as described next.
Let $\C=\{x\subseteq C\}$ be the partition of $C$ (i.e., the set of all $2^c$ subsets of $C$).
The price schedule $p_s$ of vendor $s\in\sellerset$ is a mapping from $\N^c\times\C$ to $\RR$ (the set of nonnegative real numbers), such that, for $n\in\N^c$ and $x\in\C$, $p_s(n,x)$ is the price for the bundle of products $x$ offered by $s$ under demand $n$. Let $p_s(n,\emptyset)=0$ for each $s$ and $n$.
We require that $p_s(m,x)\le p_s(n,x)$ for all $x\in\C$ if $m\ge n$ component-wise.
Letting $e_k$ be the unit vector with the $k$-th component equal to one and all other components equal to zero, we refer to $p_s^k=p_s(e_k,\{k\})$ as the \emph{base price} of item $k$ offered by $s$.
The price paid by $b$ under allocation $\mu$ is determined as follows. For each $s\in\sellerset$, let $x_b(s)=\{k\in C:\mu^k(b)=s \}$ be the set of items $b$ purchases from $s$.
Recalling that $n(s)$ denotes the demand vector of $s$ under the allocation $\mu$,
$$
\mktpriceb{b}{\mu} = \sum_{s\in\sellerset} p_s(n(s),x_b(s)).
$$
Buyers $b$ such that $\mktpriceb{b}{\mu}<\sum_{k\in C}p_{\mu^k(b)}^k$ might be willing to pay prices larger than the market price.

In general, incentive compatibility does not hold in the setting considered in this work, as buyers might benefit from misreporting their product valuations.
For example, consider a SWM allocation $\mu$ and a buyer $b$ with negative surplus $\sigma_b(\mu)$. Let $v$ be $b$'s \emph{true} valuation of the products she is matched to. If $b$ reports a valuation of $v'=v-x$, for $x>0$ such that $\mu$ remains a SWM allocation under the untruthful reporting, then she can receive a higher subsidy of $-\sigma_b(\mu)+x$ (Corollary~\ref{cor:existence} guarantees the existence of rational and stabilizing prices).
We leave this issue to future research.

Our definition of stability assumes that buyers pay base prices after deviation.
This corresponds to a situation in which deviating buyers cannot enjoy discounts.
Lu and Boutlier~\cite{luboutlier} proposed other notions of stability for scenarios in which buyers can enjoy discounts after deviation: strong stability, according to which buyers  know both demands volumes and discount schedules of all vendors,
and myopic stability, in which buyers know which discounts are triggered.
Despite efficiency coexists with these stronger notions of stability in the case of a market with a single product considered in~\cite{luboutlier},
it is an open question whether our results for multiple-item markets extend to these notions of stability.
The notion of stability of Definition~\ref{def:stability} is suitable for a setting in which prices are determined by buyers' choices (i.e., an allocation) and buyers cannot enjoy discounts by signing up to a new discount offer after deviation.

% PROOFS

\section{Proof of Proposition~\ref{prop:p-consistency}}
\label{app:p-consistency}
Let $\buyerset_1=\{b\in\buyerset:\Delta p_b>0\}$
and $\buyerset_2=\{b\in\buyerset:\Delta p_b<0\}$
be the sets of buyers that pay and receive a subsidy, respectively.
We have that
$$
\sum_{b\in\buyerset_1} \Delta p_b = -\sum_{b\in\buyerset_2} \Delta p_b.
$$
Let $B_1=\vert \buyerset_1\vert$ and $B_2=\vert \buyerset_2\vert$ be their cardinality.
We proceed by induction.
If either $B_1=1$ or $B_2=1$, then transfer $t$ can be built in a straightforward way.
Let $B_1>1$ and $B_2>1$, and assume 
$\buyerset_1=\{b_1,\ldots,b_{B_1}\}$ and $\buyerset_2=\{b'_1,\ldots,b'_{B_2}\}$.
Let
$$
k=\max\left\{j:\sum_{i=j}^{B_1}\Delta p_{b_i} \ge - \Delta p_{b'_{B_2}} \right\}.
$$
We have that $k\ge 1$.
We let \begin{equation*}
\left\lbrace
\begin{array}{ll}
\tra{b_i}{b_{B_2}}=\Delta p_{b_i} 	&,\forall i\in\{k+1,\ldots,B_1\} \\
\tra{b_k}{b_{B_2}}= - \Delta p_{b'_{B_2}} - \sum_{i=k+1}^{B_1}\Delta p_{b_i} &,i=k.
\end{array}
\right.
\end{equation*}
The defined transfer make sure that $b'_{B_2}$ receives all necessary subsidy $-\Delta p_{b'_{B_2}}$.
In addition, $b_{k+1},\ldots,b_{B_1}$ pay all their subsidy to $b'_{B_2}$.
$b_k$ has residual subsidy if $\Delta p_{b_k}> -\Delta p_{b'_{B_2}} - \sum_{i=k+1}^{B_1}\Delta p_{b_i}$.

Therefore, we reduced the problem to an equivalent problem with 
sets $\buyerset_1'$ and $\buyerset_2'$
of cardinality $B_1'\in\{k-1,k\}\le B_1$ and $B_2'=B_2-1$, respectively.
The proof of the inductive step (and therefore of the first claim) follows by observing that we can similarly reduce the cardinality of $\buyerset_1$ by one.

The second claim follows from Definition~\ref{def:p-consistency}.

%\section{Emptiness of the core}
%\label{app:example}
%Consider a market with two product types $A$ and $B$, vendors $\sellerset=\{s_1,s_2,s_3\}$, buyers $\buyerset=\{b_1,b_2,b_3\}$, and valuations
%\begin{equation*}
%\begin{array}{llll}
%b_1: 	& \vAB{b_1}{s_1}{s_1}=8, & \vAB{b_1}{s_2}{s_2}=1, & \vAB{b_1}{s_3}{s_3}=1,\\
%b_2: 	& \vAB{b_2}{s_2}{s_2}=8, & \vAB{b_2}{s_3}{s_3}=1, & \vAB{b_2}{s_1}{s_1}=1,\\
%b_3: 	& \vAB{b_3}{s_3}{s_3}=8, & \vAB{b_3}{s_1}{s_1}=1, & \vAB{b_3}{s_2}{s_2}=1,\\
%\end{array}
%\end{equation*}
%and $\vAB{b}{s}{s'}=0$ for each $b$ and  $s\neq s'$.
%Assume that $\pA{s}=\pB{s}=3$ for each $s\in\sellerset$.
%Each $s\in\sellerset$ activates a discounted price of $\pAB{s}=2$ when thresholds $\tA{s}=\tB{s}=2$ on the demand of $A$ and $B$ are met. There are three SWM allocations, symmetric with respect to each other. Consider one of them, for example $\mu(b_1)=\mu(b_2)=(s_1,s_1)$ and $\mu(b_3)=(s_3,s_3)$, with individual utilities $u_{b_1}(\mu)=6$, $u_{b_2}(\mu)=-1$, $u_{b_3}(\mu)=2$.
%Since $u_{b_1}(\mu) + u_{b_2}(\mu)=5$, regardless of the prices $p$, one of them will have a net utility of at most $2.5$.
%If $u_{b_1}(\mu,p)\le 2.5$, then $b_1$ and $b_3$ can profitably deviate by agreeing on purchasing both items from $s_3$, receiving a cumulative utility of $5$, and allocating for example $2.9$ units to $b_1$ and $2.1$ units to $b_3$.
%If instead $u_{b_1}(\mu,p)> 2.5$, then $b_2$ and $b_3$ can profitably deviate by agreeing on purchasing  from $s_2$.
%The analysis for all other SWM allocations is similar.

\section{Maximizing the social welfare is not necessary for stability}
\label{app:example2}
Consider a variation of the example in Section~\ref{app:example} above, in which $b_2$'s valuations are given by
\begin{equation*}
\begin{array}{llll}
b_2: 	& \vAB{b_2}{s_2}{s_2}=8, & \vAB{b_2}{s_3}{s_3}=1, & \vAB{b_2}{s_1}{s_1}=0.5,\\
\end{array}
\end{equation*}
The allocation $\mu$ such that $\mu(b_1)=\mu(b_2)=(s_1,s_1)$ and $\mu(b_3)=(s_3,s_3)$ has $SW(\mu)=13/2$ and is not SWM (the allocation $\mu'$ such that $\mu'(b_2)=\mu'(b_3)=(s_2,s_2)$ and $\mu'(b_1)=(s_1,s_1)$ has $SW(\mu)=7$). However, a transfer of $15/4$ from $b_1$ to $b_2$ makes $\mu$ stable.

\section{Proof of Lemma~\ref{lem:DAG}}
\label{app_lem:DAG}
The proof proceeds by induction.
First we assume that $G(\bar t)$ contains a cycle of length two, that is edges $(s_1,s_2)$ and $(s_2,s_1)$ for $s_1,s_2\in \sellerset$. We show that there exist equivalent group transfers $\bar t'$ such that either $G(\bar t') = G(\bar t) - \{(s_1,s_2)\}$ or $G(\bar t') = G(\bar t) - \{(s_2,s_1)\}$ or $G(\bar t') = G(\bar t) - \{(s_1,s_2),(s_2,s_1)\}$.
Then, we assume that the shortest cycles in $G(\bar t)$ have length $K>2$ and consider such a cycle $\K=s_1,\ldots,s_K,s_{K+1}$ (with $s_1=s_{K+1}$).
We show that there exist equivalent group transfers $\bar t'$ such that $G(\bar t')$ has a cycle of length $K-1$ obtained by replacing two adjacent edges of $\K$ with a single edge. This completes the proof as, by iterating the argument, each cycle can be reduced to a length-two cycle and finally to a single edge.

Assume $G(\bar t)$ contains edges $(s_1,s_2)$ and $(s_2,s_1)$.
Let
\begin{align*}
\X_1 & = \{ x\subseteq\sellerset: s_1\notin x, s_2\in x, \gtraset{s_1}{x} >0\},\\
\X_2 & = \{ x\subseteq\sellerset: s_2\notin x, s_1\in x, \gtraset{s_2}{x} >0\}.
\end{align*}
Let
\begin{align*}
t_{s_1} & = \sum_{x\in\X_1} \gtraset{s_1}{x},\\
t_{s_2} & = \sum_{x\in\X_2} \gtraset{s_2}{x}
\end{align*}
be respectively the total amount of cross-transfer that buyers $\posi{s_1}$ pay to all buyers $\negaset{x},x\in\X_1$ and that buyers $\posi{s_2}$ pay to buyers $\negaset{x},x\in\X_2$.

Suppose that $t_{s_1}\le t_{s_2}$. We define equivalent group transfers $\bar t'$ such that 
\begin{align}
\label{eq:dag1}
\gtraset{s_1}{x}' &= 0\quad\text{for each } x\in\X_1,
\end{align}
where buyers $\posi{s_1}$ switch a cumulative amount of transfer $t_{s_1}$ from buyers $\negaset{x},x\in\X_1$ to buyers $\negaset{x},x\in\X_2$, that is,
\begin{align}s
\label{eq:dag2}
\sum_{x\in\X_2}\gtraset{s_1}{x}' = t_{s_1} + \sum_{x\in\X_2}\gtraset{s_1}{x}.
\end{align}
Each group $\negaset{x},x\in\X_1$ receives the missing amount of transfer from buyers $\posi{s_2}$,
\begin{align}
\label{eq:dag3}
\gtraset{s_2}{x}' &= \gtraset{s_2}{x} + \gtraset{s_1}{x}\quad\text{for each } x\in\X_1,
\end{align}
for a total of $t_{s_1}$.
Buyers $\posi{s_2}$ decrease the cross-transfer to buyers $\negaset{x},x\in\X_2$ by total amount $t_{s_1}$,
\begin{align}
\label{eq:dag4}
\sum_{x\in\X_2}\gtraset{s_2}{x}' =  t_{s_2} - t_{s_1}.
\end{align}
The existence of equivalent group transfers $\bar t'$ such that~\eqref{eq:dag1}-\eqref{eq:dag4} hold is straightforward.
Observe that $\gtraset{s_1}{x}'=0$ for all $x\in\X_1$, and therefore $(s_1,s_2)\notin G(\bar t')$.
If $t_{s_2} - t_{s_1}>0$ then $\gtraset{s_2}{x}'>0$ for some $x\in\X_2$ and $(s_1,s_2)\in G(\bar t')$, otherwise $(s_1,s_2)\notin G(\bar t')$.
The proof in the case of $t_{s_1}> t_{s_2}$ similarly follows.

Assume now that the shortest cycles in $G(\bar t)$ have length $K>2$, and let $\K$ be a shortest cycle.
That is, $\K$ is formed by edges $(s_i,s_{i+1})$ for $i=1,\ldots,K$, with $s_{K+1}=s_1$.
For each $k=1,\ldots,K$ let
\begin{align*}
\X_k  &= \{ x\subseteq\sellerset: s_k\notin x, s_{k+1}\in x, \gtraset{s_k}{x} >0\},\\
t_{s_k} &= \sum_{x\in\X_k} \gtraset{s_k}{x}.
\end{align*}
Without loss of generality, assume that $s_1\in\arg\min_{s_k\in\K} t_{s_k}$, that is $t_{s_1}\le t_{s_k}$ for all $k=2,\ldots,K$ (which is always true up to node relabeling).
By the assumption that $\K$ is a cycle of minimum length, there is no chord in $G(\bar t')$, that is  $(s_k,s_j)\notin G(\bar t')$ if $s_k,s_j\in\K,s_j\neq s_{k+1}$.
We build group transfers $\bar t'$ which are equivalent to $\bar t$ and such that $(s_1,s_2)\notin G(\bar t')$ and  
$(s_i,s_{i+1})\in G(\bar t')$ for $i=2,\ldots,K$ with $s_{K+1}=s_2$ is a cycle of length $K-1$ in $G(\bar t')$.

Group transfers $\bar t'$ are defined such that
\begin{align}
\label{eq:dag5}
\gtraset{s_1}{x}' &= 0\quad\text{for each } x\in\X_1,
\end{align}
and buyers $\posi{s_1}$ switch a cumulative amount of transfer $t_{s_1}$ from buyers $\negaset{x},x\in\X_1$ to buyers $\negaset{x},x\in\X_K$,
\begin{align}
\label{eq:dag6}
\sum_{x\in\X_K}\gtraset{s_1}{x}' = t_{s_1} + \sum_{x\in\X_K}\gtraset{s_1}{x}.
\end{align}
Each group $\negaset{x},x\in\X_1$ receives the missing amount of transfer from buyers $\posi{s_K}$,
\begin{align}
\label{eq:dag7}
\gtraset{s_K}{x}' &= \gtraset{s_K}{x} + \gtraset{s_1}{x}\quad\text{for each } x\in\X_1,
\end{align}
for a total of $t_{s_1}$.
Buyers $\posi{s_K}$ decrease the cross-transfer to buyers $\negaset{x},x\in\X_K$ by total amount $t_{s_1}$,
\begin{align}
\label{eq:dag8}
\sum_{x\in\X_K}\gtraset{s_K}{x}' =  t_{s_K} - t_{s_1}.
\end{align}
The existence of equivalent group transfers $\bar t'$ such that~\eqref{eq:dag5}-\eqref{eq:dag8} hold is straightforward.
Observe that $\gtraset{s_1}{x}'=0$ for all $x\in\X_1$, and therefore $(s_1,s_2)\notin G(\bar t')$.
Buyers $\posi{s_K}$ pay a transfer of $t_{s_1}$ to  groups $\negaset{x},x\in\X_1$. This last contribution is a cross-transfer as $s_K\notin x, s_2\in x$ for each $x\in\X_1$ because $(s_1,s_K)\notin G(\bar t$). Therefore $(s_K,s_2)\in G(\bar t'$).
Moreover, if $t_{s_K} - t_{s_1}>0$ then $\gtraset{s_K}{x}'>0$ for some $x\in\X_K$ and $(s_K,s_1)\in G(\bar t')$, otherwise $(s_K,s_1)\notin G(\bar t')$.
This completes the proof.

\section{Proof of Proposition~\ref{prop:flow2}}
\label{app:propflow2}
It is straightforward to see that $\omega$ is a bijection, so we only prove the second part of the claim.
Let $\bar t=\omega(f^*)$. $\bar t$ are rational group transfers (as  $\omega$ is a bijection). Suppose by contradiction that $\bar t$ is not stabilizing, that is, condition~\eqref{eq:condition} does not hold for $\bar t$. Recall that condition~\eqref{eq:condition} reads as
\begin{equation*}
\left\lbrace
\begin{array}{ll}
\tposi{s} \ge \sum_{x:s\in x}\gtraset{s}{x} 	&\forall s\in\sellerset \\
\tnegaset{x} = \sum_{s\in x}\gtraset{s}{x} 	&\forall x\subseteq\sellerset\\
\gtraset{s}{x} = 0							&s\notin x.
\end{array}
\right.
\end{equation*}
First, suppose that $\tposi{s} < \sum_{x:s\in x}\gtraset{s}{x}$ for some $s\in\sellerset$. This would imply that the flow entering node $u_s$ is larger than the capacity of the edge $(u_s,t)$, generating a contradiction with the feasibility of the maximum flow $f^*$.
Second, suppose that $\tnegaset{x} > \sum_{s\in x}\gtraset{s}{x}$ for some $x\subseteq\sellerset,\vert x\vert \le c$.
This would imply that every flow $f'$ on $\network$ is smaller than $\sum_{x}\tnegaset{x}$, and therefore there exist no group transfers $\bar t'$ such that $\tnegaset{x} = \sum_{s\in x}\gtraset{s}{x} \forall x\subseteq\sellerset$ for all $x\subseteq\sellerset$, generating a contradiction with Theorem~\ref{thm:existence} (as feasible flows and rational group transfers are in one-to-one correspondence).
Rationality of $\bar t$ implies that $\gtraset{s}{x} = 0$ if $s\notin x$.

\section{Computational complexity for determining SWM allocations}
\label{app:expo}
We have that
$\vert\feasible\vert = {N+M^c-1\choose M^c-1}.$
To prove this, observe that computing $\vert\feasible\vert$ is equivalent to counting the number of ways in which $N$ (indistinguishable) balls can be distributed among a sorted list of $M^c$ set. Consider a line with $N+M^c-1$ empty positions. There are ${N+M^c-1\choose M^c-1}$ ways to place $M^c-1$ stones on the available positions. The occupied positions (in ascending order) represent the boundaries between the $M^c$ sets, and the cardinality of each set is the number of empty positions between two successive stones (if the first position is occupied by a stone, then the first set is empty; if the $\ell$-th and $(\ell+1)$-th positions are both occupied, then the $(\ell+1)$-th set is empty).

Using Stirling's approximation $n! \backsim (n/e)^n(2\pi n)^{1/2}$, considering $M$ constant, we have that
$$
\vert\feasible\vert \backsim \left( \frac{N}{M^c+1}+1 \right)^{M^c+1} \left( \frac{M^c+1}{N} +1 \right)^{N}  \left( \frac{1}{2\pi N} + \frac{1}{2\pi (M^c-1)}  \right)^{1/2}.
$$
Considering $M$ as a constant, we need time $\bigTheta(N^2M^c \vert\feasible\vert)$, that is,
$$
\bigTheta\left( N^2M^c \left( \frac{N}{M^c-1} \right)^{M^c-1}\right),
$$
By the upper bound ${n\choose k}\le(en/k)^k$, the time to compute a SWM allocation is
$$
\bigTheta\left( N^2M^c \left( \frac{eN}{M^c-1} \right)^{M^c-1}\right),
$$
dominated by a term $N^{M^c}$.

% Bibliography

%\bibliographystyle{plain}
%\bibliographystyle{apalike}
\bibliographystyle{plainnat}
\bibliography{discount_bib_simple}{}

\end{document}